\def\draftversion{false}

\RequirePackage{ifthen}
\ifthenelse{\equal{\draftversion}{true}}{
  \documentclass[citeautoscript,floatfix,aps,prx,twocolumn,preprintnumbers,
      superscriptaddress,longbibliography,amsmath,amssymb]{revtex4-1}
      \usepackage{showlabels}
}{
  \documentclass[citeautoscript,floatfix,aps,prx,twocolumn,
      superscriptaddress,longbibliography]{revtex4-1}
      
}

\usepackage{amsmath}
\usepackage{graphicx}
\usepackage{epstopdf}
\usepackage{natbib}
\usepackage{array}
\usepackage{dcolumn}
\usepackage{bm}

\usepackage{soul}  

\usepackage[usenames,dvipsnames]{color}

\soulregister\cite7

\ifthenelse{\equal{\draftversion}{true}}{
  \marginparwidth 2.7in
  \marginparsep 0.5in
  \newcounter{comm} 
  \def\commnext{\stepcounter{comm}}
  \def\commtext{{\bf\color{blue}[\arabic{comm}]}}
  \def\commmar{{\bf\color{blue}[\arabic{comm}]}}
  \def\dvm#1{\commnext\marginpar{\small DV\commmar: #1}\commtext}
  \def\cdm#1{\commnext\marginpar{\small CED\commmar: #1}\commtext}
  \def\msm#1{\commnext\marginpar{\small MS\commmar: #1}\commtext}
  \def\asm#1{\commnext\marginpar{\small AS\commmar: #1}\commtext}
  \def\miq#1{\commnext\marginpar{\small MR\commmar: #1}\commtext}
  \def\mlab#1{\marginpar{\small\bf #1}}
  
}{
  \def\dvm#1{}
  \def\cdm#1{}
  \def\msm#1{}
  \def\asm#1{}
  \def\miq#1{}
  \def\mlab#1{}
  
}

\def\fxc{f_{\rm xc}}
\def\Phisr{\bar{\Phi}^{\rm sr}}
\def\Philr{\bar{\Phi}^{\rm lr}}
\def\nusr{\nu_{\rm sr}}
\def\nulr{\nu_{\rm lr}}
\def\chiip{\chi_0}
\def\chiirr{\chi_{\rm ir}}
\def\chisr{\chi_{\rm sr}}
\def\chired{\chi} 
\def\wsr{W_{\rm sr}}
\def\wlr{W_{\rm lr}}
\def\rhosr{\rho^{\rm sr}}
\def\rhoscr{\rho}

\def\telr{\tilde{\epsilon}_{\rm lr}}
\def\tnulr{\tilde{\nu}_{\rm lr}}

\def\trhosr{\tilde{\rho}^{\rm sr}}
\def\tchisr{\tilde{\chi}_{\rm sr}}

\begin{document}

\title{Exact long-range dielectric screening and interatomic force constants in quasi-2D crystals}

\author{Miquel Royo}
\affiliation{Institut de Ci\`encia de Materials de Barcelona 
(ICMAB-CSIC), Campus UAB, 08193 Bellaterra, Spain}
\email{mroyo@icmab.es}

\author{Massimiliano Stengel}
\affiliation{Institut de Ci\`encia de Materials de Barcelona 
(ICMAB-CSIC), Campus UAB, 08193 Bellaterra, Spain}
\affiliation{ICREA - Instituci\'o Catalana de Recerca i Estudis Avan\c{c}ats, 08010 Barcelona, Spain}
\email{mstengel@icmab.es}

\date{\today}

\begin{abstract}
We develop a fundamental theory of the long-range electrostatic 
interactions in two-dimensional crystals by performing a rigorous study of 
the nonanalyticities of the Coulomb kernel. 
We find that the dielectric functions are best represented by 
$2\times 2$ matrices, with nonuniform macroscopic potentials 
that are two-component hyperbolic functions of the out-of-plane coordinate, $z$. 
We demonstrate our arguments by deriving
the long-range 
interatomic forces in the adiabatic regime, where we identify a formerly overlooked
dipolar coupling involving the out-of-plane components of the dynamical charges.
The resulting formula is exact up to an arbitrary multipolar order, which we
illustrate in practice via the explicit inclusion of dynamical quadrupoles.
By performing numerical tests on monolayer BN, SnS$_2$ and BaTiO$_3$ membranes, 
we show that our method allows for a drastic improvement in the description of the 
long-range electrostatic interactions, with comparable benefits to the quality of the 
interpolated phonon band structure.
\end{abstract}

\maketitle

\section{Introduction}

The separation of the interatomic force constants into short-range and long-range 
contributions has been a mainstay of lattice dynamics theory since the early 50s.~\cite{born/huang}
The work of Cochran and Cowley~\cite{cochran/cowley} has established the correct form of the long-range part
in the generic case of an anisotropic three-dimensional (3D) crystal, generalizing the earlier 
point-charge models.
The treatment, however, remained phenomenological until the seminal
work of Pick, Cohen and Martin~\cite{rmm_thesis}, where an analogous 
formula was derived in the context of first-principles theory,
and the acoustic sum rule was formally demonstrated.
The advantages of a rigorous derivation are numerous: on one hand, it paved 
the way for modern first-principles lattice dynamics, within the framework of
density-functional perturbation theory;~\cite{zein-84,baroni-87,Gonze-95,Gonze-95a,gonze,gonze-97,Baroni-01} 
on the other hand, it set the stage for further developments in linear-response 
methods, including higher-order generalizations of the Cochran-Cowley 
formula.~\cite{artlin,royo-20}

The interest in lattice-dynamical properties of two-dimensional (2D) crystals has only started 
relatively recently. As a consequence, in spite of the remarkable progress of the
past few years, the corresponding theoretical methods are not as mature as in the 3D case.
To understand the nature of the problem (i.e., why traditional algorithms 
run into trouble in 2D), consider an insulating 2D crystal suspended in vacuum.
A phonon propagating at some in-plane wavevector, ${\bf q}$,
produces stray fields that decay asymptotically as $e^{-q|z|}$, 
where $q=|{\bf q}|$ and $z$ is the out-of-plane coordinate. 
This means that, for a small enough $q$, the macroscopic electrostatic 
potential perturbation spreads over a region of space that is much larger 
than the physical thickness of the material.
Such a behaviour complicates the simulation of optical phonons
in periodic boundary conditions, as the spurious interaction between repeated
images leads to a physically incorrect description of the long-wavelength
limit unless special precautions are taken.

To address this issue, the Coulomb cutoff technique~\cite{sohrab-06,rozzi-06} 
is now routinely used in first-principles calculations of phonons~\cite{sohier-17} 
and related linear-response properties of suspended 2D systems. 
Such a treatment cures the pitfalls of a
naive supercell-based calculation, and restores the correct
physics in the small-$q$ limit by removing the undesired cross-talk between
periodically repeated images.
For example, the Coulomb cutoff nicely reproduces~\cite{sohier-17} the physically correct~\cite{deluca-20}
behavior of longitudinal (LO) optical phonons, which are degenerate in frequency 
with the corresponding transverse (TO) modes right at the Brillouin zone center, and disperse
linearly with $q$ in a vicinity of it.~\cite{sanchezportal-02,michel-11}

While the methods to perform the electronic-structure calculations are 
under control, however, the theory of the long-range electrostatic
interactions in two-dimensional crystals is still incomplete.
Their fundamental understanding is crucial for the accurate interpolation
of phonon bandstructures~\cite{sohier-17,royo-20} and electron-phonon matrix 
elements;~\cite{sohier-16,Sohier-18,Li-19,Ma-20,Ponce2020,deng2020}
to model the interaction of individual layers with the dielectric
environment (e.g., in layered heterostructures~\cite{andersen-15,mohn-18,sohier-21,sponza-20}) 
and/or experimental probes;
and to compute a number of important material properties that require a 
careful treatment of the electrostatics in the long-wavelength limit,
such as flexoelectricity.~\cite{artlin,artgr,chapter-15,Royo2019,springolo-20}
Similar issues arise in the context of electron-electron interactions,~\cite{Ando-82,Kotov-12}
electronic excitations~\cite{freysoldt-08,cudazzo-11,Berkelbach-13,Latini-15,Kylanpaa-15,huser-13}
and plasmonics~\cite{Ando-82,Yan-11,Andersen-14,Ghosh-17,agarwal-18}.
Only partial solutions where reported so far, by fitting the 
\emph{ab initio} results to 
dielectric models,~\cite{sohier-17} 
where oftentimes 
a strict 2D limit was assumed~\cite{cudazzo-11},
or with the finite thickness of the real crystal heuristically accounted 
for.~\cite{freysoldt-08,sohier-16,sohier-17,Latini-15,trolle-17} 
Systematic improvement of these models, e.g., along the lines of Ref.~\onlinecite{royo-20},
appears difficult unless a fundamental first-principles theory of the long-range
interactions in \emph{quasi-2D} crystals (that is, by explicitly treating the finite 
physical thickness of the material) is established. 

Generalizing the approach of Ref.~\onlinecite{rmm_thesis}
to the two-dimensional case, however, does not appear as an easy task.
In a quasi-2D system the electrostatic interactions are much more complex 
to understand and describe than in 3D, due to the extreme anisotropy 
of the physics between the (extended) in-plane and (microscopic)
out-of-plane directions. 
For instance, the usual tenet of 3D electrostatics of
representing the macroscopic scalar potentials via structureless 
plane waves appears inappropriate to the quasi-2D case,
where the exponential decay of stray fields in vacuum makes
the problem inherently nonuniform along $z$.
This implies, in the language of Ref.~\onlinecite{rmm_thesis}, that 
the nonanalyticities of the Coulomb kernel in 2D are not simply 
restricted to the ``head'' of the inverse dielectric matrix,
but concern an entire \emph{column} of reciprocal-space vectors 
spanning the out-of-plane direction.
Thus, separating long-range from short-range interactions, 
is \emph{per se} a highly nontrivial issue in 2D, even
at the level of the bare kernel (i.e., not considering the
additional complications related to screening).

Here we solve the aforementioned challenges by introducing a 
number of key conceptual and methodological advances.
First, we establish a rigorous and general separation between 
short-range and long-range electrostatic interactions in 2D,
both by studying the nonanalytic properties of the Coulomb kernel,
and via a physically more intuitive image-charge method.
As a direct consequence of such range separation, 
the macroscopic electrostatic potentials in 2D emerge
as two-component hyperbolic functions [$\cosh(qz)$ and $\sinh(qz)$] 
of the out-of-plane direction, $z$, reflecting 
the nonuniform nature of the long-range 
electrostatic fields.
Remarkably, the Dyson equation for the screened Coulomb interaction 
reduces then to a linear-algebra problem involving $2 \times 2$ 
matrices, i.e., is only marginally more complex than the scalar 
($1\times 1$) inverse dielectric function that is characteristic of 
the 3D case.
This result allows for a natural separation of the long-range
electrostatic potentials into even and 
odd components with respect to $z \rightarrow -z$ reflection, and
provides a unified description of both the intralayer couplings, as 
well as the interaction with external sources.
The application of our formalism to the lattice-dynamical problem
recovers the results of the existing dielectric models, 
but clearly goes beyond them, by (i) identifying a formerly overlooked 
contribution, 
i.e., the interaction between dipoles that are normal to the layer plane;
by (ii) generalizing the theory to the next lowest order in ${\bf q}$ 
via incorporation of the dynamical quadrupole tensor;~\cite{Royo2019}
and by (iii) allowing for a more accurate description of the dielectric
screening function.
Finally, we demonstrate via extensive numerical tests on 
BN, SnS$_2$ and BaTiO$_3$ membranes that our formalism allows for a 
significant and \emph{systematic} improvement in the existing methods 
for the theoretical study of phonons in 2D materials. Such an improvement 
comes at no additional cost from the computational perspective, and only 
requires a very minor addition to the existing codes.

This work is organized as follows. In Section \ref{sec:screen} and \ref{sec:pcm} 
we introduce the basic concept of range separation in the context of the 3D dielectric
matrix formalism of Pick, Cohen and Martin~\cite{rmm_thesis} (PCM). 
In Section \ref{sec:kernel2d} we present our main
conceptual achievement, which consists in identifying 
the nonanalytic part of the Coulomb kernel in quasi-2D
systems via an intuitive image-charge construction, 
and writing it as a $2\times 2$ small-space operator.
In Section \ref{sec:hyperbolic} we discuss the 
physical significance of the hyperbolic basis functions
that we use to represent the long-range Coulomb interactions.
In Section \ref{sec:multipole} we use these results to establish 
an exact formula for the long-range part of the force-constants matrix,
and relate the materials-specific
parameters to the Born effective charges, macroscopic dielectric
tensor and dynamical quadrupoles as calculated 
within modern DFPT codes; the resulting Eq.~(\ref{ddscr}) 
is another central achievement of this work.
In Section \ref{sec:range} we discuss the dependence of
many useful quantities on the range-separation parameter, 
and its implications for a physically sound description of
the dielectric function.
The remainder of this work (Sec.~\ref{sec:results}) is dedicated to the numerical 
implementation and tests of the formalism, and specifically of its performance in
the Fourier interpolation of phonon bands.

\section{Theory}

\label{sec:theory}

\subsection{Range separation of the Coulomb interactions}

\label{sec:screen}

{\em Basic definitions.} Within the adiabatic approximation, the screened Coulomb interaction $W$
links the screened potential, $V$, to an external charge perturbation,
$\rho^{\rm ext}$, as
\begin{equation}
\label{vscr}
V({\bf r}) = \int d^3r' W({\bf r},{\bf r}') \rho^{\rm ext}({\bf r'}).
\end{equation}
$W$ is, in turn, defined in terms of the bare Coulomb kernel, $\nu$,
and the irreducible polarizability $\chiirr$,
\begin{equation}
W = 
 (1- \nu \chiirr)^{-1} \nu = (1 + \nu \chi) \nu = \epsilon^{-1} \nu,
\end{equation}
where $\epsilon= 1- \nu \chiirr$ is the dielectric matrix. 
[The unity operator is a Dirac delta, $\delta({\bf r-r'})$, in
the real-space representation; it becomes a Kronecker delta 
over the reciprocal vectors in Fourier space.]
$\chiirr$ 
linearly relates the (induced) charge response of the interacting
electron system
to the screened potential;
within density-functional approaches it contains
the effects of the exchange and correlation kernel, $\fxc$,
and can be defined in terms of the independent-particle 
polarizability ($\chiip$) via a Dyson equation,
\begin{equation}
\chiirr = \chiip + \chiip \fxc \chiirr = \chiip ( 1 - \fxc \chiip)^{-1}. 
\end{equation}
By further incorporating dielectric screening effects 
we obtain the reducible polarizability, $\chi$,
\begin{equation}
\chi = \chiirr ( 1 - \nu \chiirr)^{-1} = \chiirr ( 1 + \nu \chi) = \chiirr \epsilon^{-1}.
\end{equation}

\begin{figure}
\begin{center}
\includegraphics[width=2.8in]{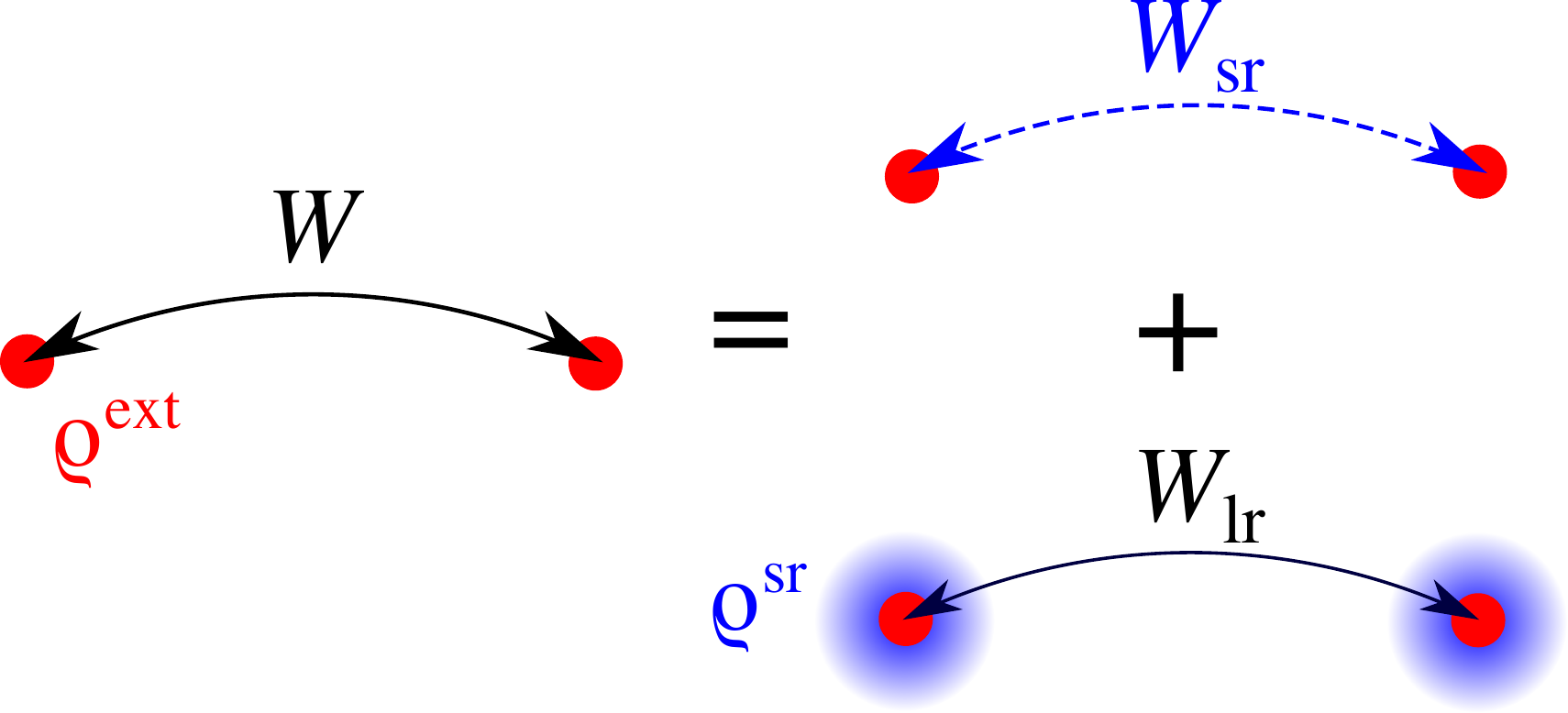}
\caption{\label{wsketch} Decomposition of the screened Coulomb interaction according to 
Eq.~(\ref{w_decomp}). The full interaction ($W$) between bare external charges ($\rho^{\rm ext}$)
is split into their mutual short-range interaction ($W_{\rm sr}$) plus the
long-range interaction ($W_{\rm lr}$) between ``dressed'' charges ($\rho^{\rm sr}$).}
\end{center}
\end{figure}

{\em Range separation.} The conceptual basis of our method consists in separating 
the bare Coulomb kernel into a short-range (SR) and a remainder long-range (LR)
part,
\begin{equation}
\nu = \nusr +  \nulr.
\label{nu_decomp}
\end{equation}
We shall assume that $\nusr$ decays exponentially in real space 
or, equivalently, can be written as an analytic function of the wavevector 
${\bf q}$ in reciprocal space; the nonanalytic part of $\nu$ is therefore
contained in $\nulr$.
We can then define a screened short-range Coulomb interaction,
\begin{equation}
\wsr = (1 - \nusr \chiirr)^{-1} \nusr = (1 + \nusr \chisr) \nusr,
\end{equation}
and similarly an intermediate polarizability function, $\chisr$, where the
electrons interact via the exchange-correlation and short-range part of 
the Coulomb kernel,
\begin{equation}
\begin{split}
\chisr =& \chiip + \chiip (\fxc + \nusr ) \chisr \\
       =& \chiirr + \chiirr \nusr \chisr 
       = \chiirr (1-\nusr \chiirr)^{-1}.   
\end{split}
\end{equation}
The operator $\epsilon_{\rm sr} = 1-\nusr \chiirr$ is a 
\emph{short-range} dielectric matrix, connecting the screened to
the external potential at the $\nusr + \fxc$ level of interaction.
We next define the screened long-range interaction as
\begin{equation}
\wlr = (1 - \nulr  \chisr)^{-1} \nulr = (1 + \nulr  \chi) \nulr,
\label{wlr}
\end{equation}
where $\epsilon_{\rm lr} = 1-\nulr \chisr$ can be regarded 
as a long-range dielectric matrix.
Based on the above ingredients, one can show (a proof is provided in Appendix~\ref{app:derivations}) 
that the following relationship holds,
\begin{equation}
\label{w_decomp}
W = \epsilon_{\rm sr}^{-1} \, \wlr \, (\epsilon_{\rm sr}^{-1})^\dagger + \wsr.
\end{equation}
This is the main formal result of this Section; an illustration of the idea 
is provided in Fig.~\ref{wsketch}.
As we shall see shortly, Eq.~(\ref{w_decomp}) constitutes
a generalization of the PCM approach, and recovers the latter as 
a special case.

{\em Lattice dynamics.} To see how this strategy works in the specific context 
of lattice dynamics, we shall combine the above results with 
the dielectric matrix formalism established in Ref.~\cite{rmm_thesis}. 
Consider a collective displacement of the sublattice $\kappa$ along $\alpha$
of the type
\begin{equation}
\label{phonon}
u^l_{\kappa \alpha} = \tau^{\bf q}_{\kappa \alpha} e^{i{\bf q\cdot R}_l}, 
\end{equation}
where $l$ is a cell index and ${\bf R}_l$ span the real-space 
Bravais lattice.
PCM's formula for the dynamical matrix at a given point ${\bf q}$ in
the Brillouin zone then reads, in our notation, as

\begin{equation}
\label{tinv}
\Phi^{\bf q}_{\kappa \alpha,\kappa' \beta} = 
 \bar{\Phi}^{\bf q}_{\kappa \alpha,\kappa' \beta} 
 - \delta_{\kappa \kappa'} \sum_{\kappa''} \bar{\Phi}^{{\bf q}=0}_{\kappa \alpha,\kappa'' \beta}.
\end{equation}
($\kappa$, $\kappa'$ and $\kappa''$ 
are basis indices, $\alpha$ and $\beta$ are Cartesian directions.) 
The matrix
\begin{equation}
\bar{\Phi}_{\kappa \alpha,\kappa' \beta} = 
   \langle \rho^{\rm ext}_{\kappa \alpha}  | \, W \, 
  | \rho^{\rm ext}_{\kappa' \beta} \rangle
\label{phiscr}
\end{equation}
describes the bare nuclear interaction screened by the total
dielectric function of the electrons at some wavevector 
${\bf q}$, which we omit from now on to avoid overburdening 
the notation.
The operator $W$ acts on the cell-periodic part of the ``external''
charge density, represented here as bra/kets. 
The latter, $|\rho^{\rm ext}_{\kappa \alpha}   \rangle$,
corresponds (see Appendix~\ref{app:derivations}) 
to the point dipoles that are induced by the
nuclear displacement pattern of Eq.~(\ref{phonon}).

The decomposition of the Coulomb kernel, Eq.~(\ref{nu_decomp}), naturally leads via Eq.~(\ref{w_decomp}) 
to a similar partition of 
$\bar{\Phi}$ 
(and hence of the force-constants matrix, $\Phi_{\kappa \alpha,\kappa' \beta}$),
\begin{equation}
\bar{\Phi} = \Phisr + \Philr,
\label{phi_decomp}
\end{equation}
where the short-range and long-range contributions are constructed according to Fig.~\ref{wsketch},
\begin{subequations}
\label{phisrlr}
\begin{align}
\Phisr_{\kappa \alpha,\kappa' \beta} &= \langle \rho^{\rm ext}_{\kappa \alpha} | \, \wsr \, | \rho^{\rm ext}_{\kappa' \beta} \rangle, \label{phisr} \\
\Philr_{\kappa \alpha,\kappa' \beta} &= 
\langle \rhosr_{\kappa \alpha} | \, \wlr \, | \rhosr_{\kappa' \beta} \rangle, \label{philr}
\end{align}
\end{subequations}
Here $\rhosr_{\kappa \alpha}$ is the ``dressed'' charge-density response 
to an atomic displacement as calculated within the SR electrostatics, 
\begin{equation}
| \rhosr_{\kappa \alpha} \rangle = 
        (1 + \chisr \nusr) \, | \rho^{\rm ext}_{\kappa \alpha} \rangle.  
\end{equation}
The round bracket corresponds to the transpose of $\epsilon_{\rm sr}^{-1}$,
which provides the formal connection between Eq.~(\ref{phi_decomp}) and Eq.~(\ref{w_decomp}).

{\em Small-space representation.} 
Thus far, we have not made any specific assumption about $\nusr$ and $\nulr$, 
except that they sum up to $\nu$.
For the practical advantages of Eq.~(\ref{phisrlr}) to become clear,
it is necessary that $\nulr$ enjoy a separable representation on a small set of basis functions, 
\begin{equation}
\label{nu_small}
\nulr = 
  \sum_{l,m=1}^{N} | \varphi_l \rangle \tilde{\nu}_{\rm lr}^{(lm)} \langle \varphi_m |,
\end{equation}
We assume that the basis functions $|\varphi_l\rangle$
have an analytic dependence on ${\bf q}$, and are smooth on the
scale of the interatomic spacings, consistent with their macroscopic 
character.
This allows to express $\Philr$ at any wavevector ${\bf q}$ 
of the Brillouin zone as a ``small space'' (see Ref.~\onlinecite{martin_book2}, Chapter 7) 
linear-algebra problem of dimension $N$ (we use a tilde to distinguish small-space from 
full-space operators), 
\begin{equation}
\Philr_{\kappa \alpha,\kappa' \beta} =  \tilde{\rho}^{\rm sr \, *}_{\kappa \alpha} \cdot \tilde{W}_{\rm lr} 
  \cdot \tilde{\rho}^{\rm sr}_{\kappa' \beta}, 
\label{philrs}
\end{equation}
where the long-range screened Coulomb interaction enjoys an analogous expression as in
the full space, 
\begin{equation}
\tilde{W}_{\rm lr} = (\underbrace{1 - \tilde{\nu}_{\rm lr}  \tilde{\chi}_{\rm sr}}_{\tilde{\epsilon}_{\rm lr}} )^{-1} 
       \tilde{\nu}_{\rm lr}.
\label{wlrs}
\end{equation}
The small-space operator $\tilde{\epsilon}_{\rm lr}$ 
acquires the physical meaning of a macroscopic dielectric matrix,
and the material-dependent ingredients entering Eq.~(\ref{philrs}) and~(\ref{wlrs}) are
defined via projections on the basis functions,
\begin{equation}
\label{rhoS} 
\tilde{\rho}^{{\rm sr},(l)}_{\kappa \alpha} =  \langle \varphi_l | \rhosr_{\kappa \alpha} \rangle, \qquad
\tilde{\chi}_{\rm sr}^{(lm)} =  \langle  \varphi_l | \, \chisr  \, | \varphi_m \rangle.
\end{equation} 
These two quantities then provide a full description of the long-range 
electrostatics in the system.

Note that the above results can be easily 
applied to the range-separation of the 
scattering potential, $V_{\kappa \alpha}({\bf r})$, of interest in 
electron-phonon problems. 
By combining Eq.~(\ref{vscr}), (\ref{w_decomp}) and (\ref{nu_small}), we find 
\begin{equation}
\label{eph}
V_{\kappa \alpha}({\bf r}) = 
 \sum_{l,m=1}^N \varphi^{\rm sr}_l({\bf r}) \tilde{W}^{(lm)}_{\rm lr} \tilde{\rho}^{{\rm sr}}_{\kappa \alpha} 
 + V_{\kappa \alpha}^{\rm sr}({\bf r}),
\end{equation}
where $V_{\kappa \alpha}^{\rm sr}$ and $\varphi^{\rm sr}_l$
are the screened potential in response 
to the phonon [Eq.~(\ref{phonon})] and to the external potential $\varphi_l$ [Eq.~(\ref{vext}) below], 
respectively, at the SR level of interaction.
Both $V_{\kappa \alpha}^{\rm sr}$ and $\varphi^{\rm sr}_l$ are, again, analytic functions
of ${\bf q}$, which allows for their efficient interpolation over the Brillouin zone;
they are available at no cost as by-product of the linear-response calculations that are 
required for the calculation of the dynamical matrix.

\begin{table}
\begin{center}
\begin{tabular}{c|c c c|c}
\hline \hline
               & \,  $\varphi-\varphi$ \,  &  \, $\varphi-\tau$  \, 
               & \, $\tau-\tau$  \, &  SCF kernel  \\
\hline
noninteracting & $\chiip$  &           &          &  $-$            \\
irreducible    & $\chiirr$ &           &          & $\fxc$          \\
short-range    & $\chisr$  & $\rhosr$  & $\Phisr$ & $\fxc + \nusr$  \\
screened       & $\chired$ & $\rhoscr$ & $\Phi$   & $\fxc + \nu$    \\
\hline \hline
\end{tabular}
\caption{ \label{respfn} Summary of the main response functions that we shall consider in
this work, together with the SCF kernel that governs the 
electron-electron interactions in each case. 
The three central columns refer to the charge response to a scalar 
potential ($\chi$), the charge response to a phonon ($\rho$), 
or the atomic forces induced by a phonon ($\Phi$).
[They can all be expressed as second derivatives of the energy with respect to
scalar potential ($\varphi$) and/or phonon ($\tau$) perturbations.]
$f_{\rm xc}$ is the exchange and correlation kernel; $\nu$ is the Coulomb kernel;
for the meaning of the short-range (sr) label, see text.
}
\end{center}
\end{table}

{\em Practical issues.} In the framework of DFPT, the main response 
functions discussed in the above paragraphs (see Table~\ref{respfn} for a summary)
can be recast as the second-order variation of the energy 
with respect to external parameters.
The force-constants matrix, for instance, involves two phonon
perturbations as defined in Eq.~(\ref{phonon}).~\cite{gonze/lee,Baroni-01}
The additional material properties that we have
introduced in the above paragraphs can be computed by 
defining $N$ new perturbations of the 
type 
\begin{equation}
\label{vext}
V^{\rm ext}({\bf r}) = \varphi_l^{\bf q} \, \varphi_l({\bf q},{\bf r}) e^{i{\bf q\cdot r}}.
\end{equation}
For the small-space polarizability ($\tilde{\chi}$) and 
charge-density response to a phonon ($\tilde{\rho}$) we have, then,
\begin{equation}
\label{respv}
\tilde{\rho}^{(l)}_{\kappa \alpha} =  
  \frac{1}{ \Omega^{[d]} } \frac{\partial^2 E}{\partial \varphi_l^{{\bf q}*}  \partial \tau^{{\bf q}}_{\kappa\alpha}}, \quad
\tilde{\chi}^{(lm)} =  \frac{1}{ \Omega^{[d]} } \frac{\partial^2 E}{\partial \varphi_l^{{\bf q}*}  \partial \varphi_m^{{\bf q}}},
\end{equation}
where $\Omega^{[d]}$ is 
the volume of the $d$-dimensional primitive cell.

The various ``flavors'' of each response function (irreducible,
screened, etc.) are determined by the type of self-consistent 
(SCF) kernel that is used in the iterative solution of the 
linear-response problem (right column in Table~\ref{respfn}).
This is particularly convenient, as it avoids the need for 
explicitly solving the Dyson equations that govern dielectric
screening at the microscopic level.
Moreover, DFPT methods allow for a more straightforward
incorporation of pseudopotentials, which are awkward to treat in the
context of the dielectric matrix formalism [e.g., in Eq.~(\ref{phiscr})
and~(\ref{phisr}) the first-order nuclear potential is that of a point 
dipole, which implies an all-electron framework].

Crucially, both ingredients entering $\Philr$, $\tilde{\rho}^{{\rm sr}}_{\kappa \alpha}$ and 
$\tilde{\chi}_{\rm sr}$, are analytic functions of ${\bf q}$, due to 
the assumed analyticity of $\nusr$.
This property is key in the perspective 
of an efficient and physically appealing
representation of $\Philr$, which can be achieved
in two different ways:
\begin{itemize}
\item One explicitly calculates  
$\tilde{\rho}^{{\rm sr}}_{\kappa \alpha}$ and 
$\tilde{\chi}_{\rm sr}$ via Eq.~(\ref{respv}),
together with $\Phisr$, on a regular mesh of 
${\bf q}$-points.
These functions are then Fourier-interpolated  
at an arbitrary ${\bf q}$-point (this is guaranteed to 
converge quickly with the mesh resolution due to 
their analytic character), where $\Philr$ 
and subsequently $\bar{\Phi}$ can be reconstructed
\emph{exactly} via Eq.~(\ref{philrs}) and Eq.~(\ref{phi_decomp}).

\item One seeks an \emph{approximate} analytical expression
(e.g. the dipole-dipole formula of Ref.~\cite{gonze/lee})
for $\Philr$ via a long-wave expansion of 
both $\tilde{\rho}^{{\rm sr}}_{\kappa \alpha}$ and 
$\tilde{\chi}_{\rm sr}$ (which is, again, 
allowed due to their analyticity)
in a vicinity of the zone center.
Typically, only few leading terms need to be
retained for an accurate description of the
long-range forces, and such quantities are 
straightforward to calculate within modern 
linear-response packages.
~\cite{gonze-97,Baroni-01}
\end{itemize}
In practice, we shall prefer the second option in
the context of this work, as it only requires minor 
modifications to the existing code implementations.

\subsection{The 3D case}

\label{sec:pcm}

As a first practical demonstration of our formalism, we shall now
use it to rederive the classic results of PCM, valid for 3D crystals. 
Following PCM, we define $\nulr$ as the ${\bf G}=0$ 
part of the Coulomb kernel in a vicinity of the zone center,
\begin{equation}
\nulr({\bf G+q,G'+q}) = \delta_{{\bf G} 0} \delta_{{\bf G}' 0} \frac{4\pi}{q^2}
\end{equation}
Since the LR kernel vanishes except for a single Fourier component,
the dimension of the ``small space'' is manifestly $N=1$, with 
structureless plane waves as basis functions, $\varphi({\bf r}) = 1$. 
This means that
both $\tilde{\chi}_{\rm sr}$ and $\tilde{\rho}^{\rm sr}_{\kappa \alpha}$ are 
scalar functions of the wavevector ${\bf q}$.
At the lowest order in ${\bf q}$, we have~\cite{rmm_thesis}
\begin{eqnarray}
\tilde{\chi}_{\rm sr} &=& -{\bf q} \cdot \bm{\chi}_{\rm mac} \cdot {\bf q} + \cdots, \\
\Omega \tilde{\rho}^{\rm sr}_{\kappa \alpha} &=& -i {\bf q} \cdot {\bf Z}_{\kappa \alpha} + \cdots,
\end{eqnarray}
where $\bm{\chi}_{\rm mac}$ and ${\bf Z}_{\kappa \alpha}$ are respectively 
the macroscopic dielectric susceptibility and Born effective 
charge tensors. 
(The tensorial components refer to the polarization direction; 
the dots stand for higher multipolar orders that are usually
neglected -- a detailed discussion of their significance can be
found in Refs.~\onlinecite{artlin,royo-20}.)
We have then, by using Eq.~(\ref{wlr}),
\begin{equation}
\tilde{W}_{\rm lr} = \frac{4 \pi}{q^2}\left(1  + \frac{4 \pi}{q^2}{\bf q} \cdot \bm{\chi}_{\rm mac} \cdot {\bf q} + \cdots \right)^{-1},
\end{equation}
which immediately leads, via Eq.~(\ref{philr}), to the established formula~\cite{rmm_thesis} for the
dipole-dipole interaction.

A disadvantage of the PCM method is that the separation between 
${\bf G}=0$ and ${\bf G}\neq 0$ terms is only meaningful in a neighborhood of the
zone center, and does not lend itself to a true range separation in 
real space.
Within our formalism, it is easy to fix this limitation. 
We define the long-range Coulomb kernel as
\begin{equation}
\nulr({\bf G+q,G'+q}) = 
   \delta_{\bf GG'} \frac{4\pi} {|{\bf G +q}|^2} e^{ -\frac{ |{\bf G +q}|^2 }{4 \Lambda^2} }.
\end{equation}
The remainder, $\nusr = \nu - \nulr$, is regular at all ${\bf q+G}$, and is 
therefore short-ranged for any nonzero value of $\Lambda$.
This corresponds to a range separation in real space in the following form,
\begin{equation}
\frac{1}{r} = \frac{{\rm erf}(r\Lambda)}{r} + \frac{{\rm erfc}(r \Lambda)}{r}.
\end{equation}
If $\Lambda^{-1}$ is large enough (say, much larger than 
the lattice parameter), then $\nulr$ only contains at most 
one nonzero element on the diagonal, while all other components
can be discarded. 
This leads to a scalar long-range formula which is very similar
to PCM's (in fact, they coincide at the leading order in ${\bf q}$),
\begin{equation}
\label{w3d}
\tilde{W}_{\rm lr} = \frac{4 \pi f(q)}{q^2}
 \left(1  + \frac{4 \pi f(q)}{q^2}{\bf q} \cdot \bm{\chi}_{\rm mac} \cdot {\bf q} + \cdots \right)^{-1},
\end{equation}
but contains a Gaussian \emph{range-separation function}, 
\begin{equation}
\label{f3d}
f(q) = e^{ -\frac{ q^2 }{4 \Lambda^2} }.
\end{equation}
The latter is reminiscent of
the usual Ewald summation techniques~\cite{gonze/lee} --
and indeed, the formalism that we have developed in this Section can
be regarded as a more sophisticated version of the Ewald method. 
It differs
in spirit from the established approach in that the range separation is applied 
here \emph{a priori} to the Coulomb kernel, and not \emph{a posteriori} to
the dipole-dipole expression of $\Philr$~\cite{gonze/lee}.
Interestingly, such an approach results in a macroscopic dielectric 
function [round bracket of Eq.~(\ref{w3d})] that explicitly depends on 
$\Lambda$ via $f(q)$; we shall come back to this point later on.

The remainder of this work will focus on how this technique 
can be generalized to systems with reduced dimensionality; 
in order to do this, we need to seek first of all an appropriate definition
of $\nulr$ in 2D.

\subsection{Coulomb kernel in two dimensions}

\label{sec:kernel2d}

In quasi-2D crystals, it is convenient to separate the 
total momentum into in-plane
(${\bf K}_\parallel = {\bf G}_\parallel + {\bf q}$, where
${\bf G}_\parallel$ belongs to the reciprocal-space Bravais lattice, 
and ${\bf q}$ is the wave vector of the perturbation) 
and out-of-plane ($k_z$) components.
Then the Coulomb kernel in open boundary conditions
can be conveniently written as a function of the in-plane momentum and 
out-of-plane real-space coordinate $z$,
\begin{equation}
\nu({\bf K}_\parallel,z-z') = 4\pi \int \frac{dk_z}{2\pi} \, \frac{e^{ik_z (z -z')}}{K_\parallel^2 + k_z^2} = 
  2 \pi \frac{e^{-K_\parallel|z-z'|}}{K_\parallel}.
\label{yukawa}
\end{equation}
By analogy with the 3D case, one may be tempted to
identify $\nulr$ with the ${\bf G}_\parallel =0$ 
component of Eq.~(\ref{yukawa}), which is clearly nonanalytic.
Doing so, however, would be unfit to our purposes: unlike the 
${\bf G} =0$ component of the Coulomb kernel in 3D, $\nu({\bf q},z-z')$ 
is not a scalar, but an \emph{operator} that depends nontrivially on $z$ and $z'$.
The practical appeal of the range separation method discussed
in the previous Section rests on the representability of $\nulr$ 
in a ``small'' space, where only one (as in the 3D case) or few 
physical degrees of freedom of macroscopic character are treated
explicitly.
The function $\nu({\bf q},z-z')$ clearly violates such a condition.

\begin{figure}
\begin{center}
\includegraphics[width=3.2in]{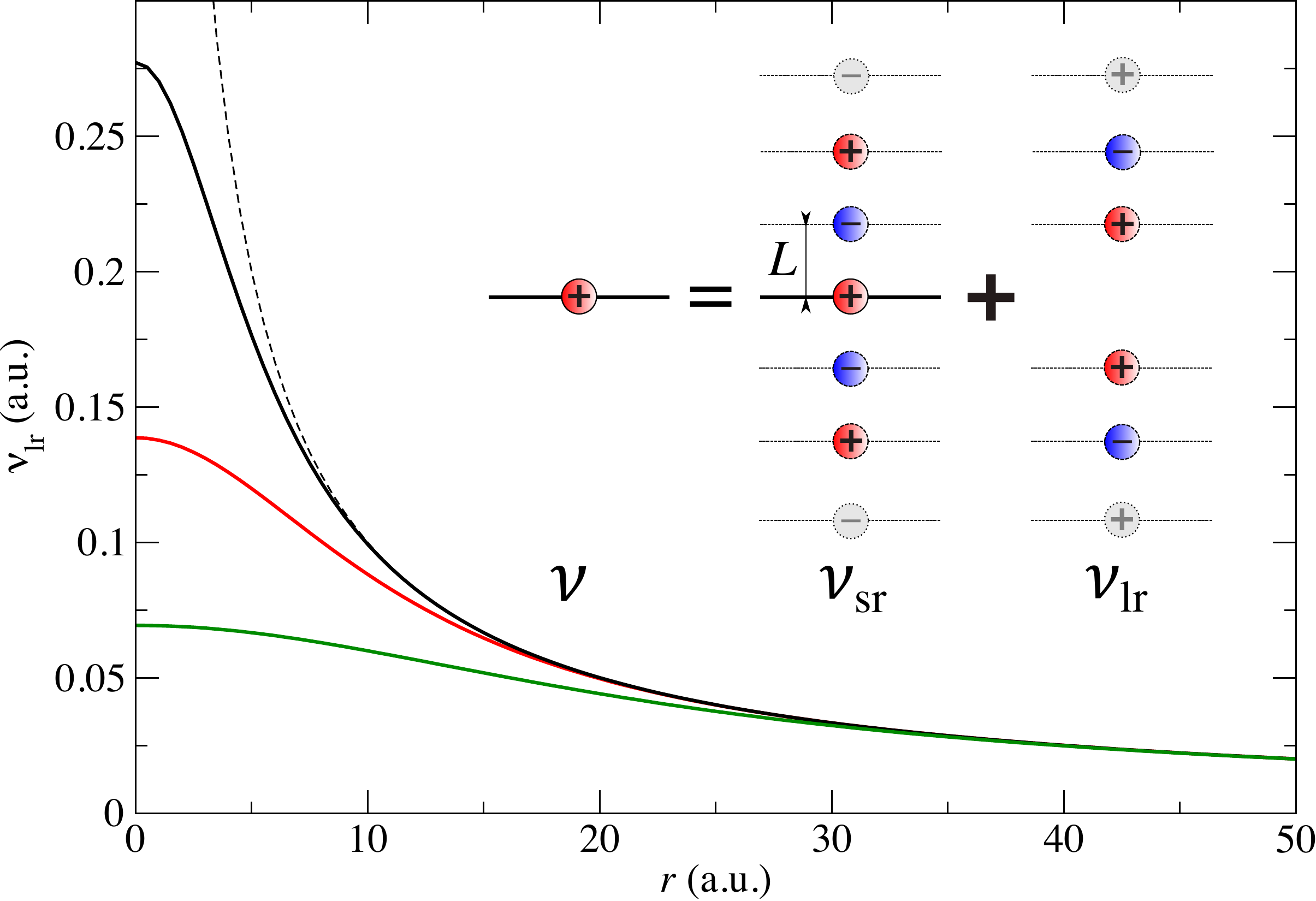}
\caption{\label{kernel} Real-space representation of the long-range 
Coulomb kernel as defined in the text. The curves correspond to 
the choices $L=5$ a.u. (black), $L=10$ a.u. (red) and $L=20$ a.u. (green).
The Coulomb potential $1/r$ is shown as a dashed curve for comparison. The 
inset provides an intuitive illustration of the decomposition of Eq.~(\ref{decomp_nu})
as an image-charge method.}
\end{center}
\end{figure}

To overcome this obstacle, and thereby achieve a sound separation between
short-range and long-range interactions, we shall use 
the image-charge construction that is illustrated in 
Fig.~\ref{kernel} (inset).
In particular, we shall define the short-range Coulomb
kernel as follows,
\begin{equation}
\nusr({\bf K}_\parallel,z-z') = \sum_n (-1)^n \nu({\bf K}_\parallel,z-z'-nL). 
\label{decomp_nu}
\end{equation}
This consists in replacing an external charge perturbation 
(represented as a red circle with a ``+'' symbol in Fig.~\ref{kernel})
with an infinite array of images, spaced by a distance $L$ along
the out-of-plane direction and taken with alternating
signs.
We assume that the parameter $L$ 
is larger than the physical thickness of the layer, in 
such a way that neighboring images of the ground-state 
electronic density have vanishing overlap.
For the same reason, we shall restrict our attention to the 
range $|z-z'|<L$, which is the physically relevant regime for 
Coulombic interactions within the layer.

Clearly, 
$\nu_{\rm sr}$ is short-ranged, 
as the electrostatic potential produced by the linear 
array of alternating point charges vanishes exponentially
for $r_\parallel \gg L$, where $r_\parallel$ is
the in-plane distance from the array.
Then, the long-range interactions 
must be entirely contained in $\nulr$, which is
defined via Eq.~(\ref{nu_decomp}) as the remainder,
$\nulr=\nu-\nusr$.
To illustrate this fact, in the main panel of 
Fig.~\ref{kernel} we show a real-space 
representation of $\nulr$ as it results from 
such a construction.
As expected, $\nulr$ deviates significantly 
from the Coulombic potential only for 
$r_\parallel< L$, where it avoids the $1/r_\parallel$ 
divergence of the latter and tends smoothly to a constant value instead.

The short-range Coulomb kernel as defined in Eq.~(\ref{decomp_nu})
appears exotic at first sight, so the fact that it has been available 
for several decades in mainstream implementations of DFPT~\cite{Giannozzi_2009,abinit} 
may come as a surprise to the reader.
In a plane-wave electronic-structure code, suspended 2D 
crystals are routinely calculated by means of the 
\emph{supercell approach}; this consists in repeating 
the system periodically along the vacuum direction, while
setting the distance between images to some 
sufficiently large value, $L$, to avoid any unphysical 
cross-talk. 
As we anticipated in the introduction, long-wavelength phonons 
are problematic to simulate within such a computational setup, as 
spurious electrostatic interactions between images cannot be avoided in 
the ${\bf q} \rightarrow 0$ limit, unless special precautions (e.g.,
by means of the Coulomb truncation method) are taken~\cite{Sohier2017a}. 
Such unphysical interactions, however, can be exploited to 
our advantage, as they provide a straightforward
first-principles implementation of Eq.~(\ref{decomp_nu}).
Indeed, a phonon traveling in the superlattice with
momentum $(q_x,q_y,q_z=\pi/L)$, i.e., located at the Brillouin-zone
boundary along the vacuum direction,
introduces a phase delay of 180$^\circ$ between neighboring 
images, which reproduces the alternating signs of our image-charge 
construction.

An explicit formula for $\nulr$ can
be derived by carrying out the summation of the
$n\neq0$ terms in Eq.~(\ref{decomp_nu}).
Due to our assumption of $|z-z'|<L$, the argument 
$z-z'-nL$ is defined negative for $n\geq 1$, and positive 
for $n\leq 1$. 
After a straightforward algebraic
manipulation (see Appendix~\ref{app:derivations}) 
we arrive then at
\begin{equation}
\nulr({\bf K}_\parallel,z-z') = \frac{2\pi f({K}_\parallel)}{{K}_\parallel} 
  \cosh [{K}_\parallel(z-z')],
\label{nulr1}
\end{equation}
where the range-separation function in the prefactor is
\begin{equation}
f(K_\parallel)=  1 - \tanh \left(\frac{K_\parallel L}{2} \right).
\label{fk}
\end{equation}
$f(K_\parallel)$ is monotonously decreasing and vanishes exponentially 
for $K_\parallel \gg 1/L$:
for a charge modulation of sufficiently short wavelength, the images 
do not ``see'' each other, as the stray fields decay faster than the
vacuum thickness. 
In such a regime, the ``zone-boundary'' electrostatics coincides with the 
correct one and $\nulr$, which is defined as the difference, vanishes.
The parameter $L$ defines the length scale of the range separation
(see the main panel of Fig.~\ref{kernel}),
and plays a similar role as the Gaussian width $\Lambda$ in
Eq.~(\ref{f3d}).
Note that $f(K_\parallel)$ has a linear behavior ($f \simeq 1 - K_\parallel L / 2$) 
for small $K_\parallel$, which contrasts with the quadratic behavior of 
its 3D counterpart; we regard this outcome as a consequence of the reduced 
dimensionality.
One can verify that Eq.~(\ref{nulr1}) exactly reproduces
the nonanalytic behavior of the full kernel, Eq.~(\ref{yukawa}),
at any order in ${\bf K}_\parallel$.

The hyperbolic cosine diverges exponentially
for large arguments, which may raise some questions about the numerical
stability of Eq.~(\ref{nulr1});
also, one may wonder how we ended up with an unbounded
potential when the original kernel of Eq.~(\ref{yukawa}) is manifestly a bounded
function of $|z-z'|$ at any nonzero $q$.
We stress that the cosh of Eq.~(\ref{nulr1}) is really intended as
a \emph{truncated} hyperbolic cosine ($\nulr$ is defined in the range
$|z-z'|<L$), in the same spirit of the 
Coulomb truncation method.~\cite{sohrab-06,rozzi-06,sohier-16}
(Our parameter $L$ corresponds to half the supercell
length within the latter approach.)
And, in fact, our definitions of $\nu_{\rm lr}$ and $\nu_{\rm sr}$,
once represented on a plane-wave basis set, exactly sum up to
the truncated Coulomb kernel as defined by Sohier {\em et al.}~\cite{sohier-16}
(a formal proof is provided in Appendix~\ref{lautrec}).
Incidentally, our derivations show that Sohier's method 
can also be understood as an image-charge construction: 
it only differs from $\nusr$ by a factor $e^{-qL}$ in
the odd-numbered terms of Eq.~(\ref{decomp_nu}).

With these results in hand, we are now ready to attack the representability issue 
that we raised at the beginning of this Section. At first sight, it might seem  
that we haven't made much progress -- Eq.~(\ref{nulr1}) is still expressed 
as a nontrivial function of $z$ and $z'$. 
By using the elementary bisection formula of the hyperbolic cosine, 
however,
one can equivalently write Eq.~(\ref{nulr1})
as
\begin{equation}
\label{nulr2}
\nulr({\bf q},z-z') =  \bm{\varphi}(z) \cdot
\tnulr({\bf q})
\cdot \bm{\varphi}(z'),
\end{equation} 
in terms of the small-space
operator
\begin{equation}
\label{nulrS}
\tnulr({\bf q}) =\frac{2\pi f(q)}{q} \, 
\begin{pmatrix}
1              &   0        \\
0              &  -1              
\end{pmatrix},
\end{equation} 
and the two-component macroscopic potential,
\begin{equation}
\bm{\varphi}(z) = \left[ \cosh (qz) , \sinh (qz) \right]. 
\label{potential}
\end{equation}
[We assume that $f(K_\parallel)\simeq 0$ for any ${\bf G}_\parallel \neq 0$,
leaving us with a simple ${\bf q}$-dependence of $\tnulr$.]
Eq.~(\ref{nulr2}) now provides the sought-after separable 
representation of the long-range Coulomb kernel.
In spite of the apparent complexity of the electrostatic problem in 2D,
with the extreme anisotropy of the physics between the in-plane and
out-of-plane directions and the consequent inhomogeneity of the 
stray fields [here reflected in the nonuniform nature of the
basis functions, Eq.~(\ref{potential})], we
have managed to represent the long-range Coulomb interactions in
a space whose dimensionality is only slightly larger than that of
the trivial 3D case; we regard this as a remarkable conceptual 
achievement of this work.

The fact that the hyperbolic basis functions diverge 
exponentially with $z$ is, again, not an issue in practice, since our main
focus is on \emph{intralayer} interactions, occurring within a 
bounded region $|z-z'|<L$. 
In the next Section we shall further corroborate their physical 
soundness by addressing the electrostatic potentials \emph{far away}
from the layer, which mediate its coupling to the dielectric environment 
and/or external probes.

\subsection{Hyperbolic functions}

\label{sec:hyperbolic}

To understand the physics that lies behind the two-component
nature of the electrostatic potentials and operators, it is useful to 
recall some basic properties of the hyperbolic functions 
appearing in Eq.~(\ref{potential}).
The hyperbolic cosine is manifestly an even function of $z$,
while the sine is odd: at the lowest order, the former reduces
to an electric field acting parallel to the plane, 
while the latter corresponds to a perpendicular field. 
(To reflect this fact, we shall indicate the two components of 
the relevant matrices and vectors with the ``$\parallel$'' and
``$\perp$'' symbols henceforth.)
This means that the cosh and sinh potentials mediate electrostatic 
interactions between charge densities that are, respectively, even and 
odd with respect to $z$-reflection. 
The emergence of a mirror-odd component marks a drastic
departure from the 3D case, where transverse electric fields
are forbidden by the translational periodicity of the crystal
Hamiltonian.
Based on the above, we can interpret the hyperbolic 
basis functions as the \emph{quasi-2D generalization of modulated
electric fields}, respectively oriented in-plane (cosh) or 
out-of-plane (sinh).
This generalization is unique, as there is a unique solution
to the Laplace equation in all space once the boundary condition 
at the $z=0$ plane is fully specified.
This also means that the cosh and sinh functions constitute
a \emph{complete} basis for expanding an arbitrary electrostatic
potential that is produced by external charges (i.e. located
outside the volume of the layer).

As a consequence, the ``small-space'' representation of the perturbed 
charge density [Eq.~(\ref{rhoS})] must be relevant to describing not only 
the long-range interactions within the layer, but also the exponentially
decaying vacuum fields \emph{outside} the layer.
To see this, consider an isolated 2D layer with a screened charge perturbation of the form
\begin{equation}
\rho^{\rm ext}({\bf r}) = e^{i{\bf q \cdot r}} \rho^{\bf q}(z),
\end{equation}
where $\rho^{\bf q}(z)$ is the planar average of the cell-periodic part. 
The electrostatic potential generated by $\rho^{\bf q}(z)$ can be written as a 
convolution in real space with the kernel of Eq.~(\ref{yukawa}),
\begin{equation}
\label{vz}
V^{\bf q}(z) = \int dz' \, \nu({\bf q},z-z') \rho^{\bf q}(z').
\end{equation}
If we consider a point that is located far
enough from the layer that the perturbed density vanishes, Eq.~(\ref{vz})
reduces to
\begin{equation}
V^{\bf q}(z) = \frac{2\pi}{q} \left\{\begin{array}{lr}
  e^{-qz} \int dz' \, e^{qz'} \rho^{\bf q}(z') & \text{for } z \gg 0, \\
  e^{qz} \int dz' \, e^{-qz'} \rho^{\bf q}(z') & \text{for } z \ll 0. 
        \end{array}\right.
\end{equation}
After observing that $e^{\pm qz} = \cosh(qz) \pm \sinh(qz)$,
we obtain, for $|z|\gg 0$,
\begin{equation} 
V^{\bf q}(z) = \frac{2\pi}{q} e^{-q|z|} \left[  \rho^\parallel({\bf q}) + {\rm sgn}(z) \rho^\perp({\bf q}) \right], \\
\end{equation}
where ${\rm sgn}(z)$ is the sign function, and we have defined the two-component 
charge-density perturbation by combining Eq.~(\ref{rhoS}) with Eq.~(\ref{potential}),
\begin{subequations}
\begin{align}
\rho^\parallel({\bf q}) =& \int dz \, \rho^{\bf q} (z) \cosh (qz), \\
\rho^\perp({\bf q}) =& \int dz \, \rho^{\bf q} (z) \sinh (qz).
\end{align}
\end{subequations}
This shows that the stray fields in the vacuum region are entirely specified
by the ``small space'' representation of the screened charge density, 
thereby further substantiating its physical significance.

\subsection{Long-range interatomic forces}

\label{sec:multipole}
In order to write the LR part of the dynamical matrix according to Eq.~(\ref{philr}) we 
shall define the small-space representations of the short-range polarizability ($\chisr$) and 
charge-density response to a phonon ($\rho_{\kappa \alpha}^{{\rm sr}}$) by using
Eq.~(\ref{potential}) in conjunction with the formalism of Sec.~\ref{sec:screen}.
Then, the observation that $\cosh(qz)$ and  $\sinh(qz)/q$ are
both analytic functions of ${\bf q}$ naturally leads to a long-wave
expansion of $\tchisr$ and $\trhosr$.
Regarding $\tchisr$, we have
\begin{equation}
\tchisr({\bf q}) = - 
\begin{pmatrix}
 {\bf q} \cdot \bm{\alpha}^\parallel \cdot {\bf q}              &   q {\bm{\beta}} \cdot {\bf q}        \\
  q {\bm{\beta}} \cdot {\bf q}             &   q^2    \alpha^{\perp}          
\end{pmatrix} + O(q^4),
\label{chi_approx}
\end{equation} 
where we have introduced the in-plane ($\bm{\alpha}^\parallel$) 
and out-of-plane ($\alpha^\perp$) macroscopic polarizabilities of the layer, and 
${\beta}_m$ denotes the off-diagonal elements that couple in-plane 
and out-of-plane dipoles;
their relation to the macroscopic dielectric tensor of the supercell is
described in Appendix~\ref{app:multipoles}.
Note that these relationships are \emph{exact}, i.e. they do 
not rely on any assumption regarding the physical properties
of the layer, unlike the dielectric model of Ref.~\onlinecite{sohier-16}.

The charge-response functions, on the other hand,
can be conveniently expanded as 
\begin{subequations} \label{rho_approx}
\begin{align}
\tilde{\rho}^{\rm sr,\parallel}_{\kappa \alpha}({\bf q})  =& -\frac{iq_\beta}{\sqrt{S}}   \underbrace{
         \left[ \hat{Z}^{(\beta)}_{\kappa \alpha} - i\frac{q_\gamma}{2}  \left(
             \hat{Q}^{(\beta\gamma)}_{\kappa \alpha} - \delta_{\beta \gamma} \hat{Q}^{(zz)}_{\kappa \alpha} \right)
                                       + \cdots \right]  }_{\mathcal{Z}^{(\beta)}_{\kappa \alpha}({\bf q}) } \nonumber \\
            & \times  e^{-i{\bf q} \cdot \bm{\tau}_\kappa}, \\
\tilde{\rho}^{\rm sr,\perp}_{\kappa \alpha}({\bf q})     =& \frac{q}{\sqrt{S}}           \underbrace{ 
     \left[  \hat{Z}^{(z)}_{\kappa \alpha}   
     -iq_\beta \hat{Q}^{(z \beta)}_{\kappa \alpha} + \cdots \right] }_{\mathcal{Z}^{\perp}_{\kappa \alpha}({\bf q}) }
     e^{-i{\bf q} \cdot \bm{\tau}_\kappa},
\end{align}
\end{subequations}
where $S$ is the cell surface [see Eq.~(\ref{fou2D})], 
the complex phase is a structure factor that depends on the in-plane location of the
atom $\kappa$ within the cell, and
we have indicated as $\hat{\bf Z}_{\kappa \alpha}$ and $\hat{\bf Q}_{\kappa \alpha}$ the 
dynamical dipole and quadrupole tensors in 2D. 
These generally differ from their
standard definitions in 3D (see Appendix~\ref{app:multipoles} for details):
(i) the electrical boundary conditions are set to short circuit in plane, and 
open circuit along $z$, consistent with the ``zone-boundary'' electrostatics;
(ii) the Cartesian moments along $z$ are calculated \emph{with respect to the
$z=0$ plane}, which corresponds to the center of the 2D layer.

The way $\hat{Q}^{(zz)}_{\kappa \alpha}$ enters
Eq.~(\ref{rho_approx}), which stems from the asymptotic
expansion of the hyperbolic cosine, $\cosh(qz) \simeq
1 + q^2 z^2 / 2$, might appear surprising at first
sight.
To see its physical significance note that, in 
classical electrostatics, only the \emph{traceless}
part of the Cartesian multipole tensor~\cite{applequist-89} 
produces long-range electrostatic fields (see Appendix~\ref{app:traceless}).
In two dimensions, this implies that the diagonal 
elements of the quadrupolar tensor only contribute
to the long-range forces via their \emph{difference},
consistent with Eq.~(\ref{rho_approx});
we regard this outcome 
as a further demonstration of the 
internal consistency of our theory.

While the above formalism is entirely general, 
for simplicity we shall focus henceforth on 2D crystals that 
enjoy a mirror plane at $z=0$.
This assumption implies that the off-diagonal component of the
polarizability, ${\beta}_m$, vanishes by symmetry, and the diagonal elements 
of the screened Coulomb interactions can be treated as two 
separate scalar problems. 
By plugging the long-wave expansions of the densities, Eq.~(\ref{rho_approx}), and
the dielectric functions, Eq.~(\ref{chi_approx}), into Eq.~(\ref{philrs}), we obtain the following formula 
for the long-range interatomic forces,
\begin{widetext}
\begin{equation}
\bar{\Phi}^{\rm lr}_{\kappa \alpha, \kappa' \beta}({\bf q}) = 
\frac{2\pi f(q)}{Sq}  \left( \frac{  ({\bf q} \cdot \bm{\mathcal{Z}})^*_{\kappa \alpha} \,
  ({\bf q} \cdot \bm{\mathcal{Z}})_{\kappa' \beta} }
 {\tilde{\epsilon}_\parallel({\bf q})} 
     - q^2 \frac{\mathcal{Z}^{\perp*}_{\kappa \alpha} \, \mathcal{Z}^\perp_{\kappa' \beta}}
    {\tilde{\epsilon}_\perp({\bf q})}
     \right)
e^{-i{\bf q} \cdot (\bm{\tau}_{\kappa'} - \bm{\tau}_\kappa)}.
\label{ddscr}
\end{equation}
\end{widetext}
where 
\begin{subequations} \label{dielectric}
\begin{align}
\tilde{\epsilon}_\parallel({\bf q}) &= 1 + \frac{2\pi f(q)}{q} {\bf q} \cdot ({\bm \alpha}^\parallel + \cdots) \cdot {\bf q}, \\
\tilde{\epsilon}_\perp    ({\bf q}) &= 1 - 2\pi q f(q)  \, (\alpha^\perp + \cdots).
\end{align}
\end{subequations}
$\tilde{\epsilon}_{\parallel}$ and $\tilde{\epsilon}_{\perp}$ refer to the 
diagonal components of the small-space dielectric matrix, $\telr$. 
At leading order in ${\bf q}$, they
correspond, respectively, to the monopolar and dipolar response functions that
were considered in earlier works~\cite{andersen-15,sohier-21};
the dots stand for the terms $O(q^4)$ and higher in Eq.~(\ref{chi_approx}).
$\mathcal{Z}$ are the dynamical dipoles, corresponding to the 
square brackets in Eq.~(\ref{rho_approx}), which generally depend on ${\bf q}$
via quadrupolar and higher-order terms.
Eq.~(\ref{ddscr}) describes the long-range electrostatic interactions \emph{exactly} up 
to an arbitrary multipolar order; this is the second central result of this work.

By truncating the expansions of Eq.~(\ref{chi_approx}) and Eq.~(\ref{rho_approx}) 
to their leading orders in ${\bf q}$, we recover an \emph{approximate} representation 
of the long-range force constants 
that can be directly compared with earlier works on the subject.
The mirror-even part of Eq.~(\ref{ddscr}) is consistent with the formula
proposed by Sohier {\em et al.}~\cite{sohier-17}, with the most obvious 
difference that the range-separation function $f(q)$ in the prefactor is replaced a 
Gaussian, $g(q) = e^{-\frac{q^2}{4\Lambda} }$, therein. 
Both functions tend to unity at ${\bf q}=0$ and may therefore appear
equivalent at first sight.
Our $f(q)$ as given by Eq.~(\ref{fk}), however, displays a linear
(rather than quadratic) dependence at small $q$, which is key
to reproducing the nonanalytic behavior of the long-range Coulomb 
kernel exactly.
Interestingly, in our formula $f(q)$ also appears in the 
definition of the dielectric function, Eq.~(\ref{dielectric});
we shall come back to this 
point in the following Section.

The mirror-odd part (second term in the round bracket) 
of Eq.~(\ref{ddscr}) is, to the best of our knowledge,
an original result of this work. (The contribution of the 
out-of-plane dipoles to the long-range potentials discussed by 
Ref.~\cite{deng2020} has mirror-even quadrupolar character, and
therefore is qualitatively different; see Appendix~\ref{app:multipoles}
for further details.)
Remarkably, the interaction between
out-of-plane dipoles enters with a \emph{negative} sign, which originates
from Eq.~(\ref{nulrS}).
To rationalize this outcome, note that
an unsupported insulating film imposes open-circuit electrical
boundary conditions
on out-of-plane dipoles, which implies that, at ${\bf q}=0$,
optical phonon modes experience a full depolarizing
field along $z$. Such physics is well described by the ``zone-boundary''
electrostatics that we discussed earlier.
When moving away from $\Gamma$, the spatial modulation of the dipole moments
acts as an effective Yukawa-like \emph{screening}, which progressively
weakens the effects of the depolarizing field; 
the physics is not dissimilar to the driving force towards domain formation 
in low-dimensional ferroelectrics.
As we shall see in the results section, this implies
that the ZO branch (optical modes with polarization out of plane)
approaches ${\bf q}=0$ with a linear dispersion, similarly to
LO modes but with a negative slope.
Remarkably, in the mirror-odd
component of the dielectric function, Eq.~(\ref{dielectric}), 
the out-of-plane polarizability of the layer also enters 
with negative sign, which implies that $\epsilon^\perp({\bf q})$
is always smaller than one. 
This outcome might bear intriguing connections to
the theory of negative capacitance~\cite{zubko-16} 
effects in thin-film ferroelectrics; we regard this as a 
fascinating topic to explore in future studies.

\subsection{The range separation parameter}

\label{sec:range}

As we have mentioned earlier, an interesting outcome of our derivations 
is that the small-space dielectric function, $\telr$, explicitly depends on the 
range-separation parameter via $f(q)$.
In particular, the $f(q)$ prefactor suppresses the polarizability
contribution at large momenta, and $\telr({\bf q})$ tends to 
unity for $q \gg L^{-1}$ (i.e., at length scales where the physics of
the dielectric screening is microscopic in character).
This behavior is common to both the 2D [Eq.~(\ref{dielectric})] and the
3D [Eq.~(\ref{w3d})] cases, i.e., it does not depend on dimensionality but
appears to be a general consequence of the formalism developed in Sec.~\ref{sec:screen}.
The appearance of a fictitious parameter ($L$ or $\Lambda$) may appear undesirable;
it is, however, a natural manifestation of the arbitrariness in the separation between 
what we regard as ``macroscopic'' and ``local field'' effects, which is inherent
to our strategy.
This issue is well known in other contexts: e.g., in the ``nanosmoothing'' 
techniques~\cite{junquera-07,baldereschi-88} that are used to extract macroscopic
physical information from microscopic first-principles data; or in the
popular Ewald method, which can be regarded as a straightforward application 
of our formalism to a system of classical point charges.

In the case of the mirror-odd component, the progressive 
suppression of the polarizability contribution for increasing $q$
is not only a direct consequence of the above arguments, 
but is also an essential ingredient for a mathematically stable 
description of the long-range interactions.
Indeed, the contribution of the layer polarizability $\alpha^\perp$ 
enters with a negative sign, which would lead to a vanishing denominator 
in Eq.~(\ref{ddscr}) if $f(q)$ were neglected (i.e., set to 
unity) in Eq.~(\ref{dielectric}).
One can show that the stability condition is 
\begin{equation}
\label{stability}
L > 4\pi \alpha_\perp.
\end{equation}
By recalling the definition of $\alpha_\perp$, Eq.~(\ref{alp_perp}), 
one quickly realizes that the above condition marks the crossover
between a positive and a negative value of $\epsilon_{zz}^{-1}$, 
the inverse dielectric constant of the hypothetical cell of thickness $L$
that we use to represent our 2D crystal.
Thus, assuming a ``strict 2D limit'' (e.g., following the
guidelines of Ref.~\onlinecite{cudazzo-11}) would be unphysical in
the context of the out-of-plane dielectric function: an infinitesimally
thin layer with a finite out-of-plane polarizability would inevitably lead
to divergencies in the screened Coulomb interaction at short distances.
In the language of Sec.~\ref{sec:screen}, one can equivalently say that 
the small-space operator $\telr$ must be an invertible matrix for our 
method to be physically sensible and mathematically stable; the above
considerations show that $L$ must be chosen wisely for this condition
to hold.

As a matter of fact, all analytic response functions that one 
calculates within the SR Coulomb kernel depend on $L$ implicitly via 
the $L$-dependence of the latter.
This raises the obvious question of whether the small-momentum expansion 
coefficients of $\trhosr$ [Eq.~(\ref{rho_approx})] and/or $\tchisr$ 
[Eq.~(\ref{chi_approx})] are affected by this issue.
One can show that the lowest orders in ${\bf q}$ of either
function, including all quantities that are explicitly 
mentioned in Eq.~(\ref{rho_approx}) and Eq.~(\ref{chi_approx}), are 
independent of $L$; their respective $L$-dependence
kicks in at the octupolar level for $\trhosr$ and at $O(q^4)$ 
for $\tchisr$.
The 2D and 3D cases are, again, qualitatively similar in these
regards: a demonstration that the quadrupolar moments are 
independent of the range-separation parameter (a fictitious 
Thomas-Fermi screening length was used) in 3D crystals, while 
octupoles are not, can be found, respectively in Ref.~\cite{martin}
and Ref.~\cite{artlin}.

Of course, the \emph{screened} counterparts of the 
charge-density response and polarizability must be independent
of $L$, consistent with their definition in free-boundary conditions.
Interestingly, we have
\begin{subequations}
\begin{align}
\tilde{\rho}_{\kappa \alpha}({\bf q}) &= \telr^{-1}({\bf q}) \, \trhosr_{\kappa \alpha}({\bf q}),   \\
\tilde{\chi}({\bf q})                 &= \telr^{-1}({\bf q}) \, \tchisr({\bf q}).
\end{align}
\label{screened}
\end{subequations}
This means that the implicit 
$L$-dependence of the SR quantities (which originates from the 
modifications to the short-range Coulomb kernel that
a variation of $L$ entails)
cancels out \emph{exactly} with an analogous dependence of $\epsilon$
when the former are divided by the latter.
Such a cancellation becomes only approximate when the multipolar
representations of both $\rhosr$ and $\epsilon$ are truncated, and
such a deviation can be used to gauge the overall accuracy of the
method.

\begin{table*}
\setlength{\tabcolsep}{8pt}
\begin{center}
\begin{tabular}{r|rr|rr|rrrrr}\hline\hline
                          &   \multicolumn{1}{c}{B}     &    \multicolumn{1}{c}{N}     &   \multicolumn{1}{c}{Sn}  &    \multicolumn{1}{c}{S(1)}      &  \multicolumn{1}{c}{Ba(1)} &  \multicolumn{1}{c}{Ti}    &  \multicolumn{1}{c}{O(1)}    &  \multicolumn{1}{c}{O(2)}    &    \multicolumn{1}{c}{O(3)}   \\  
\hline
$Z_{\kappa x }^{(x)}$   &  2.685  & $-$2.685 & 4.814 & $-$2.407  &  2.946 &  6.603 & $-$2.487 & $-$2.285 & $-$5.237  \\
$Z_{\kappa z }^{(z)}$   &  0.246  & $-$0.246 & 0.343 & $-$0.171  & 0.482 & 0.947  & $-$0.675 &  $-$0.280 & $-$0.280 \\ 
\hline
$Q_{\kappa x }^{(xy)}$  &  4.261    &  0.384   &   &      3.700 & & & & & \\ 
$Q_{\kappa y }^{(xx)}$  &  4.261    &  0.384   &   &      3.700 & & & & & \\
$Q_{\kappa y }^{(yy)}$  &  $-$4.261 & $-$0.384 &   &   $-$3.700 & & & & & \\ 
$Q_{\kappa y }^{(yz)}$  & & &   & $-$0.298  & 1.356  &  & $-$0.972  &  & \\ 
$Q_{\kappa z }^{(xx)}$  & & &   & $-$2.932  & $-$24.552 &  & 21.641 &  & \\
$Q_{\kappa z }^{(zz)}$  & & &     & 0.231  & 4.605  &  & $-$3.868  &  & \\
\hline
$\alpha_{\parallel}$    &  \multicolumn{2}{c}{1.882}   & \multicolumn{2}{c}{6.629} &  \multicolumn{5}{c}{4.461}   \\
$\alpha_{\perp}$        &  \multicolumn{2}{c}{0.310}   & \multicolumn{2}{c}{0.720} &  \multicolumn{5}{c}{0.900}  \\ 
\hline \hline
\end{tabular}
\end{center}
\caption{Cartesian components of dynamical dipole, quadrupole and polarizability tensors
(in the ``zone-boundary'' electrostatics). Atomic units are used; only linearly 
independent coefficients are shown. The numbering of the atoms refers to
the convention of Fig.~\ref{figmat}.
} 
\label{Tab_all}
\end{table*}

\section{Results}

\label{sec:results}
\begin{figure}[b!]
\begin{center}
\includegraphics[width=3.2in]{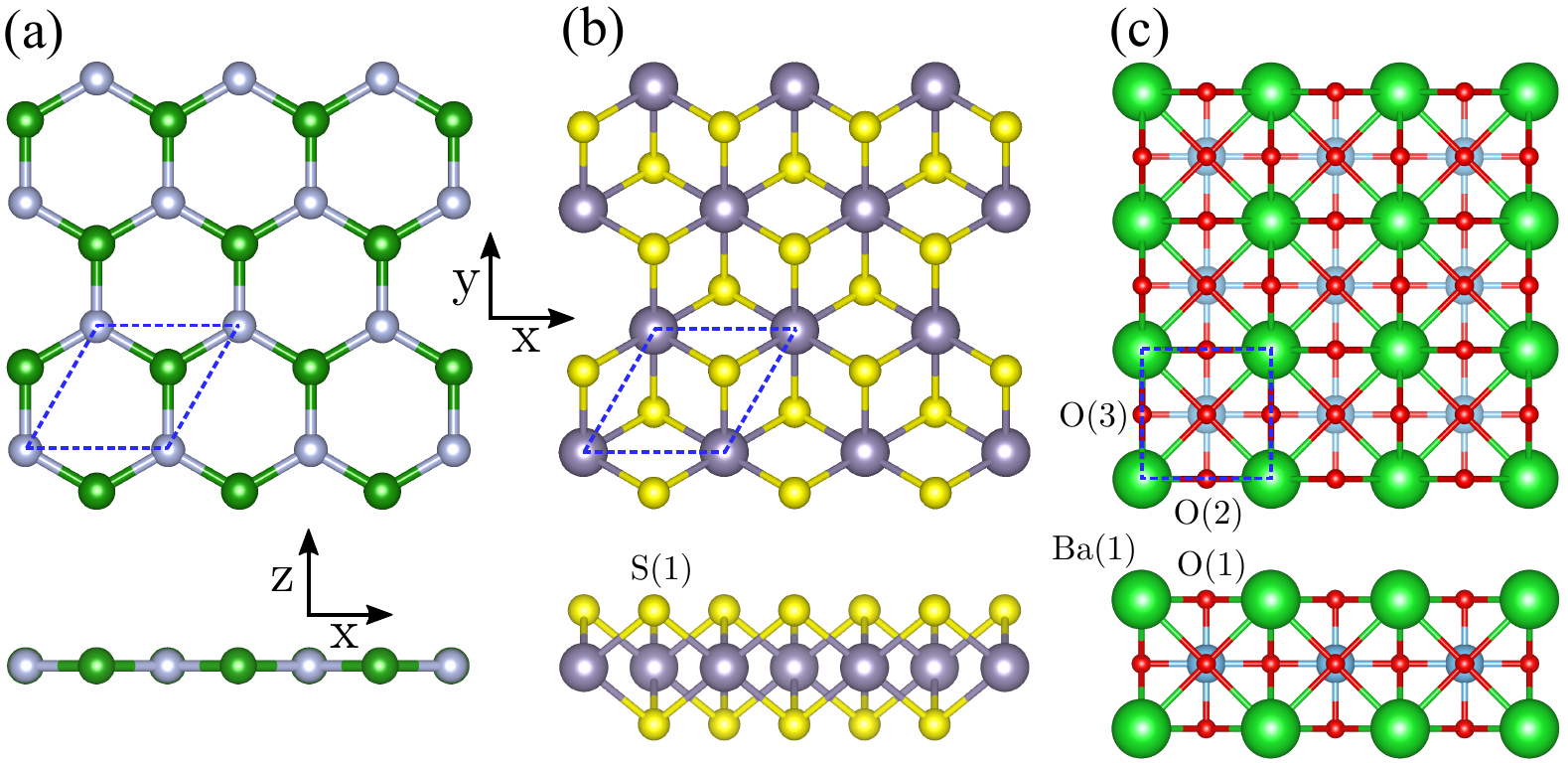}
\caption{\label{figmat} Atomic structure of the simulated 2D materials:
 BN (a), SnS$_2$ (b), BaTiO$_3$ (c).}
\end{center}
\end{figure}

We shall now benchmark the performance of our
method regarding the Fourier interpolation of
the dynamical matrix elements and eigenvalues
(phonon bands).
Our computational model consists in the three materials illustrated in 
Fig.~\ref{figmat}, i.e., in two representative 2D monolayer 
crystals, BN and SnS$_2$, and a thin membrane of BaTiO$_3$.
(The latter consists in a tetragonal stacking of three 
BaO/TiO$_2$/BaO layers, the relaxed structure being non piezoelectric.)
To start with, 
we shall present the calculated 
physical parameters for our materials set.

\subsection{Calculation of the physical constants}

\label{sec:physctns}

Our calculations are performed within the local-density approximation
as implemented in ABINIT~\cite{ABINIT2020},
by using optimized norm-conserving Vanderbilt pseudopotentials~\cite{HamannPRB13}
from the PseudoDojo.~\cite{VANSETTEN201839}
For all the materials considered, we use a plane-wave
cutoff of 80 Hartree and a $12\times12\times1$
${\bf k}$-point grid. 
The length of the supercell in the out-of-plane $z$-direction is 
set to 40 Bohr in all cases. 
Before performing the linear-response calculations, we optimize the atomic 
positions and cell parameters of the unperturbed systems to a stringent 
tolerance ($10^{-8}$ and $10^{-6}$ atomic units for residual stress
and forces, respectively).
The linear-response quantities (dielectric tensor, 
dynamical charges~\cite{gonze/lee} and quadrupoles~\cite{Royo2019}) 
necessary to build the LR dynamical matrix are then computed with the DFPT 
and longwave drivers of ABINIT~\cite{ABINIT2020} and subsequently 
transformed to the zone-boundary electrostatics following the recipe of Appendix~\ref{3Dto2D}. 
The results are reported in Table~\ref{Tab_all}.

\subsection{Interpolation of the dynamical matrix \label{DMinterp}}

We shall now test the performance
of our method regarding the Fourier-interpolation of
the phonon bands. 
In particular, we shall benchmark the results of our interpolation method, 
against the \emph{exact} DFPT phonon frequencies  
and the frequencies obtained by means of the 
phenomenological 2D Fourier interpolation of Sohier \emph{et al.}~\cite{sohier-17}
both accessible via the Quantum Espresso~\cite{Giannozzi_2009,Giannozzi_2017} (QE) suite.
In the course of our tests, we have detected a missing factor of $2\pi/a_0$
in the QE subroutine (version 6.5) that builds the long-range interactions following the 
guidelines of Ref.~\cite{sohier-17}.
Such a factor likely passed unnoticed in earlier works,~\cite{sohier-17} as it
is close to one (in atomic Bohr units) in all materials studied therein.
For a fair comparison, in the following we shall present results obtained 
\emph{after} having fixed this issue.
(Our fix will be incorporated in future releases of the 
software, presumably starting from v.6.8.)

To calculate the dynamical matrices we use the 
``Coulomb truncation'' method~\cite{sohrab-06} as implemented~\cite{sohier-16,Sohier2017a}
in the linear-response module of QE.
For consistency, we use the same computational parameters,
exchange and correlation functionals and pseudopotentials
as in our ABINIT calculations (see Sec.~\ref{sec:physctns}).
Prior to performing the actual calculations, we carefully check
the compatibility between ABINIT and Quantum Espresso calculations
by comparing the main linear-response quantities 
(polarizabilities and Born effective charges) that can be obtained 
through both packages, obtaining essentially no differences (within a 
tolerance of four significant digits).

Once the dynamical matrices are calculated on a
discrete mesh (${\bf q}_i$) of points spanning the 2D Brillouin zone
of the crystal, 
we evaluate the approximate (A) long-range interactions,
$\bar{\Phi}^{\rm lr,A}_{\kappa \alpha, \kappa' \beta}({\bf q}_i)$,
via the truncated Eq.~(\ref{ddscr}) on the same 2D mesh, and use it 
to define an approximate short-range part as
\begin{equation}
\Phi^{\rm sr,A}({\bf q}_i) = \Phi({\bf q}_i) - \Phi^{\rm lr,A}({\bf q}_i).
\label{approx_sr}
\end{equation}
[$\Phi^{\rm lr,A}$ is obtained from $\bar{\Phi}^{\rm lr,A}$ after
enforcing translational invariance via Eq.~(\ref{tinv}).]
Finally, $\Phi^{\rm sr,A}_{\kappa \alpha, \kappa' \beta}({\bf q}_i)$ is 
Fourier-interpolated to obtain the short-range dynamical matrix
at an arbitrary ${\bf q}$, and eventually the full dynamical 
matrix once the long-range part $\Phi^{\rm lr,A}_{\kappa \alpha, \kappa' \beta}({\bf q})$
is added back to it.
Again, $\bar{\Phi}^{\rm lr,A}_{\kappa \alpha, \kappa' \beta}({\bf q})$ 
only depends on $L$ (the only free parameter) via the range-separation function 
$f(q)$ that is contained in Eq.~(\ref{ddscr}) and Eq.~(\ref{dielectric}).

\begin{figure}
\centering
    \includegraphics[width=1.0\columnwidth]{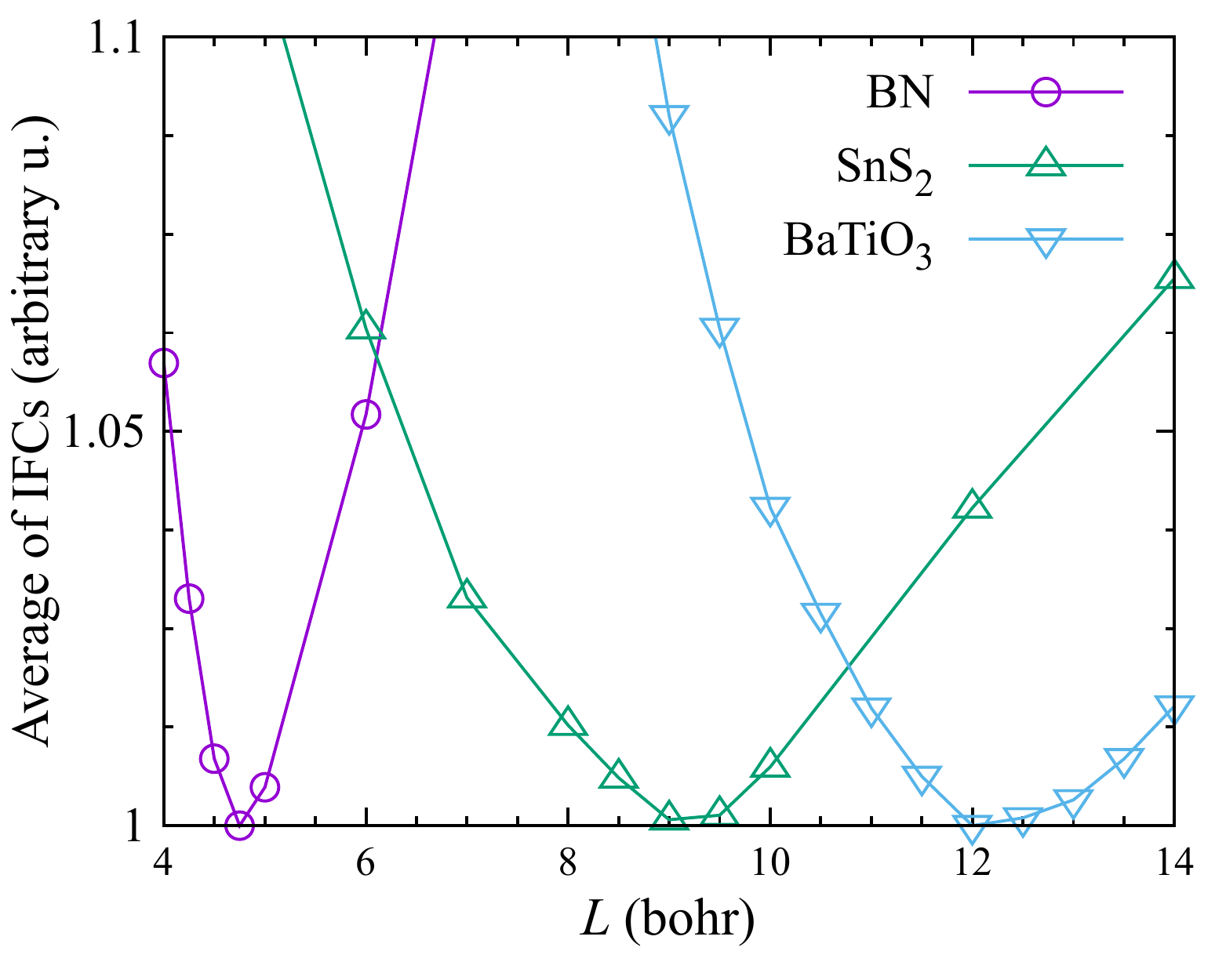}
    \caption{Average of short-range IFCs (absolute value excluding self-interactions) 
    as a function 
    of the $L$ parameter used to generate the long-range dynamical matrices. For
    each of the three materials studied in this work the y-axis data have been scaled by 
    the minimum calculated value.}
    \label{Fig_IFCs}
\end{figure}

To determine the optimal value of $L$, 
we estimate the accuracy of the interpolation at a given $L$
by requiring that the decay of the ``sr'' force constants in
real space be as fast as possible.
In practice, we define an indicator by summing
up the absolute values of the short-range IFCs in real space,
\begin{equation}
d(L) = \sum'_{\kappa \kappa' \, l} \sum_{\alpha \beta}
 |\Phi^{\rm sr,A}_{\kappa \alpha, \kappa' \beta}(0,l)|,
 \label{dl}
\end{equation}
where the prime means that self-interactions are excluded.
The minimum of $d(L)$ yields then the sought-after value of $L$.
This only entails a minimal computational burden, since 
it only requires recalculating $\bar{\Phi}^{\rm lr,A}_{\kappa \alpha, \kappa' \beta}({\bf q})$ 
several times at different values of $L$. This is done
at the level of the post-processing program (i.e., it does 
not imply running additional linear-response calculations).
The results for our tested materials are shown in Fig.~\ref{Fig_IFCs}.
For BN, we checked that the value of $L$ optimized via Eq.~(\ref{dl}) is
consistent with our conclusions based on the analysis of the 
screened charge, following the guidelines of Sec.~\ref{sec:range}.

\begin{figure*}
\centering
    \includegraphics[width=0.9\textwidth]{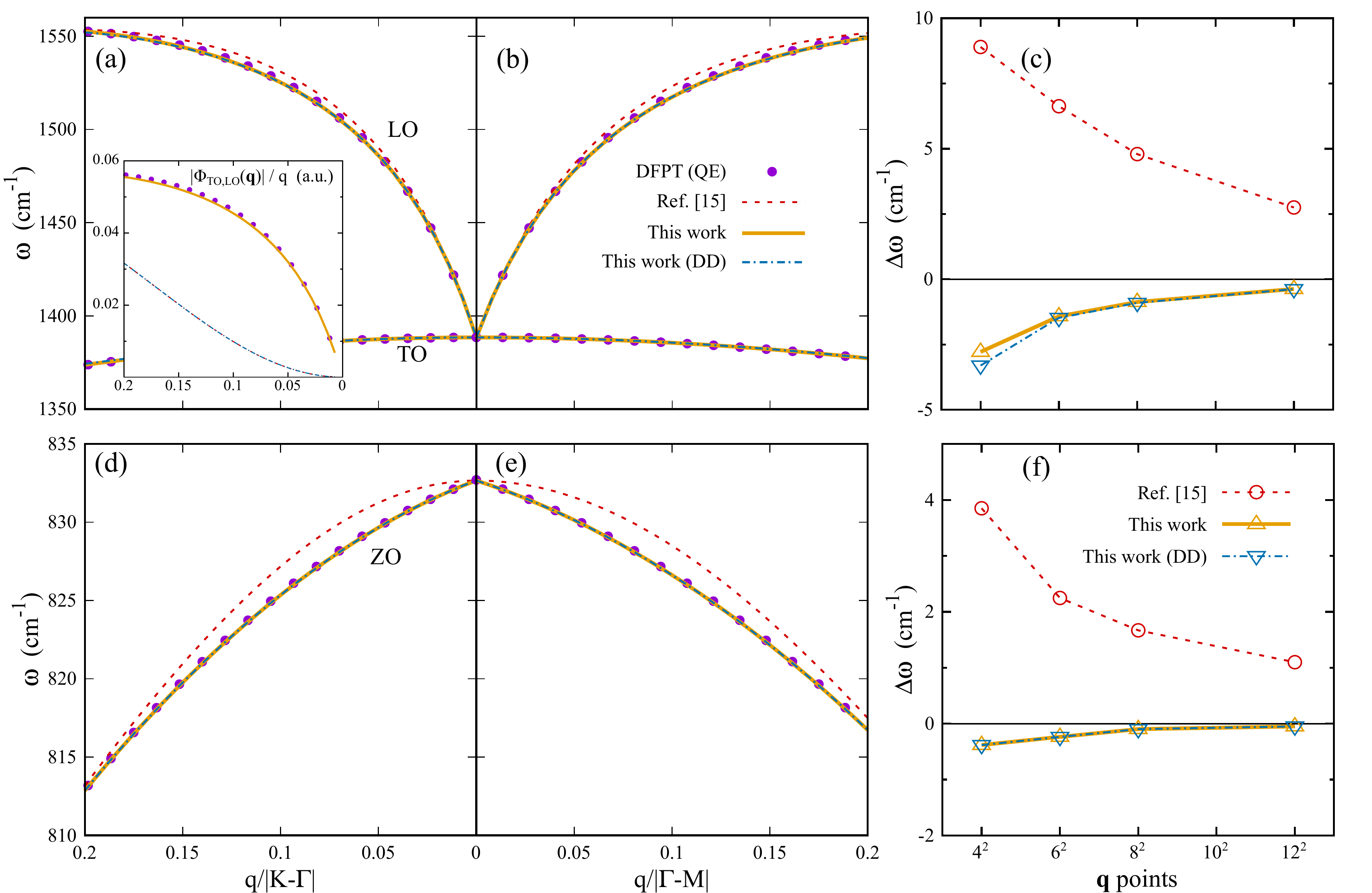}
    \caption{ {\bf BN.} Dispersion of the LO, TO (a--b) and ZO (d--e) phonon branches in the
    long-wavelength limit along the K--$\Gamma$ [(a) and (d)] and $\Gamma$--M 
    [(b) and (e)] segments in the 2D Brillouin zone. 
    The curves show the results of the 2D Fourier interpolation 
    method of Ref.~\onlinecite{sohier-17} as implemented in Quantum Espresso 
    after having corrected the bug in the implementation (see text), and of the
    method developed in this work, either with or without (DD) 
    dynamical quadrupoles; exact [DFPT (QE)] phonon frequencies are
    shown as circles.
    All interpolations are performed 
    on a grid of $8\times8$ in-plane {\bf q} points. 
    (c) and (f) show the absolute error resulting from the different
    interpolation methods as a function of the {\bf q}-mesh
    density for the LO and ZO branches, respectively. 
    DFPT frequencies are taken as 
    reference, and the error is evaluated at the {\bf q} point lying halfway
    between $\Gamma$ and the first commensurate {\bf q} point along the $\Gamma$--M
    segment.
    The inset in (a) shows the element of the $\Gamma$-modes representation of the force-constants matrix 
    [Eq.~(\ref{eq_ifcproj})] that describes the coupling between the
    TO and LO modes along the K--$\Gamma$ segment. 
    Absolute values divided by the modulus of {\bf q}   
    are represented.}
    \label{Fig_BN}
\end{figure*}

\subsubsection{BN}

We begin by applying our scheme to study the long-wavelength dispersion of the optical 
phonons in monolayer BN. 
Phonons in BN have been the subject of several works in the framework
of tight-binding models~\cite{sanchezportal-02}, classical potentials~\cite{michel-11,park-18} 
or first-principles electronic-structure theory;~\cite{wirtz-03,sohier-17}
therefore, this material constitutes an excellent first benchmark for our method.
Fig.~\ref{Fig_BN} shows the results obtained with our Fourier interpolation 
formalism and using a value of $L=4.5$ Bohr which, as anticipated in 
the previous section and confirmed by the data represented in Fig.~\ref{Fig_IFCs}, is optimal in order to 
minimize the spread of the IFCs. 
Compared with the bands obtained by following the interpolation of 
Sohier \emph{et al.},~\cite{sohier-17}
our method manifestly improves the description of both the LO and ZO branches, 
accurately reproducing the exact DFPT frequencies. 

Regarding the LO mode, 
we ascribe this improvement to our more accurate treatment of the long-range 
2D screening function, while the inclusion of dynamical quadrupoles appears
to have a minor impact on the interpolated LO frequencies.
To see this, we repeated the interpolation procedure while neglecting dynamical
quadrupoles in $\bar{\Phi}^{\rm lr,A}$ (dot-dashed blue curves in Fig.~\ref{Fig_BN}), 
obtaining negligible differences. 
For a more quantitative comparison, we show in Figs.~\ref{Fig_BN}(c)
the deviation from the exact LO branch as a function of the {\bf q}-mesh
resolution: our method is highly accurate already at a coarse $4\times4$ mesh,
while earlier treatments result in a much slower convergence.

\begin{figure*}
    \centering
    \includegraphics[width=0.9\textwidth]{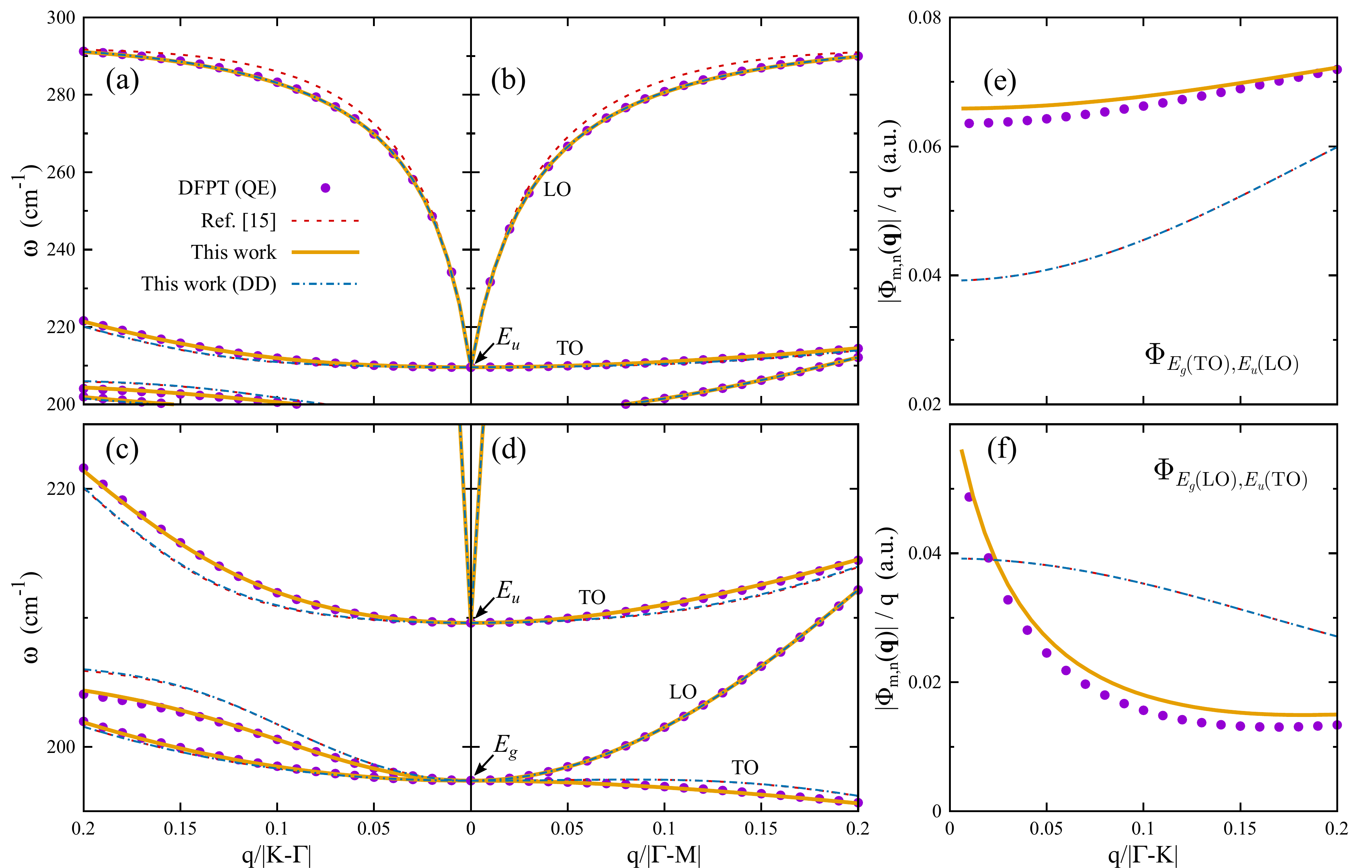}
    \caption{ {\bf SnS$_2$.} (a-b): Dispersion of the optical $E_u$ 
    branches in the
    long-wavelength limit along the K--$\Gamma$ (a) and $\Gamma$--M 
    (b) segments in the 2D Brillouin zone. 
    (c--d): Dispersion of the lower-energy $E_u$ and the two
    $E_g$ branches along the K--$\Gamma$ (c) and $\Gamma$--M 
    (d) segments.
    Interpolations are performed on a grid of $8\times8$ in-plane 
    {\bf q} points. 
    (e-f): Elements of the $\Gamma$-modes representation of the 
    force-constants matrix corresponding to the 
    $E_g$(TO)--$E_u$(LO) (e) and $E_g$(LO)--$E_u$(TO) (f) 
    couplings along the $\Gamma$--K segment. 
    Absolute values divided by $|{\bf q}|$ are shown.}
    \label{Fig_SnS2}
\end{figure*}

The seemingly negligible impact of the dynamical quadrupoles in the interpolation
of the LO frequencies is surprising, so we decided to investigate this point further.
We find that the quadrupolar terms are important to reproduce the correct 
\emph{interactions} between modes, corresponding to the off-diagonal elements
of the dynamical matrix, in the long-wavelength limit.
To illustrate this point, we project the force-constants matrix 
at a given wavevector, ${\bf q}$, onto the $\Gamma$-point mode eigenvectors 
($u^m_{\kappa\alpha}$, $m$ being a mode
index), appropriately modulated by a position-dependent complex phase,
\begin{equation}
 \Phi_{m,n}({\bf q})=\langle u^m_{\kappa\alpha} | e^{-i\,{\bf q} \cdot \bm{\tau}_\kappa} 
 \Phi_{\kappa\alpha,\kappa'\beta}({\bf q}) \, e^{i\,{\bf q} \cdot \bm{\tau}_{\kappa'}}
 |u^n_{\kappa'\beta} \rangle .
 \label{eq_ifcproj}
\end{equation}
In the inset of Fig.~\ref{Fig_BN}(a) we plot the off-diagonal element of 
$\Phi_{m,n}({\bf q})$, quantifying the strength of the coupling between the LO and TO
modes, along a portion of the K--$\Gamma$ segment.
As above, we compare the exact DFPT values with the results of the 
Fourier interpolation, which we perform either including or excluding the 
contribution of the dynamical quadrupoles.
Clearly, the quadrupoles play a crucial role in ensuring that the
long-wave limit is accurately described. Note the qualitative error
of the dipole-dipole interpolation, which approaches $\Gamma$ quadratically
instead of linearly.
As a matter of fact, the specific treatment of the dipole-dipole terms 
has no effect on the coupling between these two modes. All the models
(except that including the dynamical quadrupoles), or even a complete
neglect of the long-range interactions during the interpolation, 
yield exactly the same result [see inset of Fig.~\ref{Fig_BN}(a)].

Regarding the ZO branch of Fig.~\ref{Fig_BN}(d--e), note its characteristic 
linear dispersion when approaching the $\Gamma$ point, which is reminiscent 
of the LO branch except for the (negative) sign of the group velocity. 
This behavior, as we mentioned earlier, stems from the out-of-plane 
dipole-dipole interactions, which were neglected in earlier works.
Indeed, when such interactions are left untreated, as in the 
dashed curves of Fig.~\ref{Fig_BN}(d--e), the Fourier interpolation
results a quadratic dispersion, and a discrepancy that decays 
very slowly with the {\bf q}-mesh resolution [Fig.~\ref{Fig_BN}(f)].
Our method clearly reproduces the qualitatively correct 
physics in the long-wavelength limit, with an excellent match
between the interpolated and exact frequencies already at
the coarsest mesh resolution that we have considered [Fig.~\ref{Fig_BN}(f)].
Note that dynamical quadrupoles are irrelevant here, since their 
effect on the mirror-odd part of the electrostatics vanishes by 
symmetry.

\subsubsection{SnS$_2$}

SnS$_2$ has been the focus of several studies lately, both in its
bulk~\cite{zhen-20} and monolayer~\cite{shafique-17} forms.
Its main interest lies in the very low lattice thermal conductivity,~\cite{he-18}
which is important for thermoelectric efficiency.
Clearly, an accurate representation of phonon frequencies is
key to these applications, which motivates its consideration as
a representative testcase.
Note that in the case of SnS$_2$, a larger (compared to BN) value of 
$L=9$ bohr yields an optimally fast decay of the IFCs (see Fig.~\ref{Fig_IFCs}) 
and has been therefore used in the interpolation.

Fig.~\ref{Fig_SnS2} shows the dispersion of the four (out of six) optical branches
that are lowest in energy.
The modes of Fig.~\ref{Fig_SnS2}(a--b)
originate from the doubly degenerate $E_u$ mode at
the $\Gamma$ point, with a calculated frequency of 
$\omega(E_u)=$210 cm$^{-1}$, and correspond to the LO and TO modes
with in-plane polarization.
Similarly to the BN case, our interpolation scheme has a
most visible impact on the highest LO mode, where our improved 
treatment of screening results in an excellent match with 
the exact DFPT frequencies.
Again, the effect of quadrupoles appears to be unimportant for
the interpolation of the LO branch, which is very well described
already at the dipole-dipole level.

In Fig.~\ref{Fig_SnS2}(c--d) we show, on a magnified vertical
scale, the TO branch of the aforementioned $E_u$ doublet,
together with two additional branches deriving from the 
Raman-active $E_g$ modes [$\omega(E_g)=$197 cm$^{-1}$].
Here, contrary to the above examples, the inclusion of dynamical 
quadrupoles is important to reproduce the correct phonon dispersion. 
This can be clearly appreciated by the comparison with the results
of the lower-order models (limited to dipole-dipole interactions),
which significantly deviate from the exact DFPT frequencies.
The latter, on the other hand, are matched by the full electrostatic 
model with excellent accuracy.

To understand the reason why the dipole-dipole model is
inaccurate for these bands, we have quantified the quadrupolar
strength of each mode along the two relevant $\hat{\bf q}$-directions,
by projecting the calculated components of $Q_{\kappa \alpha}^{(\beta \gamma)}$
on the corresponding $\Gamma$-point eigenvectors.
Interestingly, the largest discrepancies between the two electrostatic
models are observed along the branches where dynamical quadrupoles
vanish by symmetry, which might appear counterintuitive at first sight.
However, one must keep in mind that Fourier interpolation is
a global operation on the 2D Brillouin zone. 
This means that residual nonanalyticities 
in the ``short-range'' dynamical matrix affect the quality of all 
interpolated branches, including those that are not directly concerned
by macroscopic electric fields.
It turns out that, similarly to the BN case, the inclusion of dynamical 
quadrupoles significantly improves the description of the \emph{off-diagonal}
matrix elements, which in SnS$_2$ couple $E_g$(TO) with $E_u$(LO) 
[Fig.~\ref{Fig_SnS2}(e)] and $E_g$(LO) with $E_u$(TO) [Fig.~\ref{Fig_SnS2}(f)] when moving 
away from $\Gamma$.
Here, the impact on the phonon frequencies is much larger than in BN 
because the relative closeness in energy of the interacting branches amplifies 
the effect. We have likewise confirmed that dipole-dipole interactions
play no role in interpolating these off-diagonal matrix elements.

Note that two additional optical branches, respectively of
$A_{g}$ and $A_{u}$ symmetry, are present at higher energies (not shown); the 
electrostatic corrections, while present, have a relatively lesser impact
on their interpolated frequencies.

\begin{figure}
    \centering
    \includegraphics[width=1.0\columnwidth]{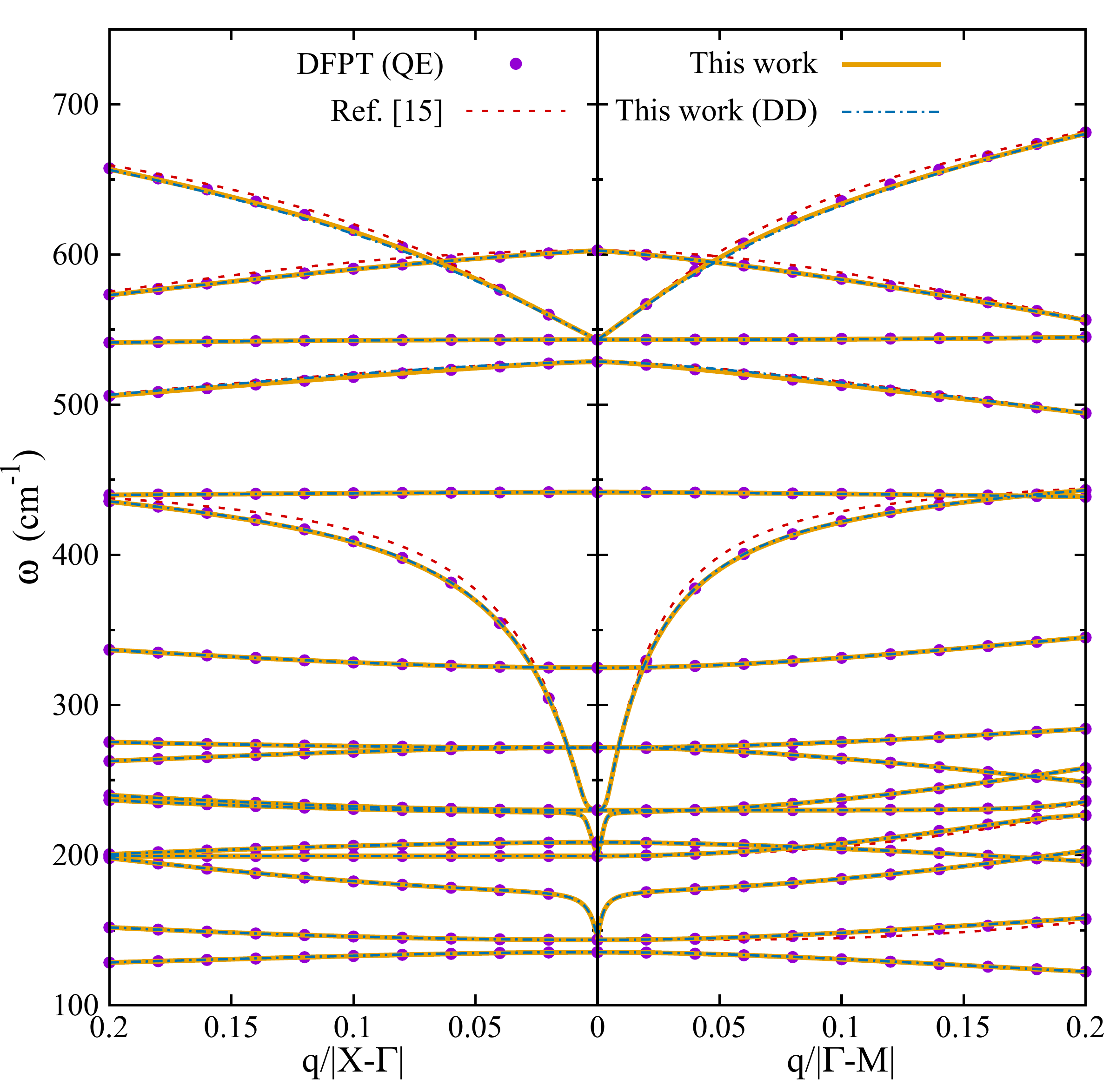}
    \caption{ {\bf BaTiO$_3$ membrane.} Dispersion of the optical phonon 
    branches in the
    long-wavelength limit along the X--$\Gamma$ and $\Gamma$--M 
    segments. The interpolated frequencies are obtained from the
    dynamical matrices calculated on a grid of $8\times8$ in-plane 
    {\bf q} points. }
    \label{Fig_BTO}
\end{figure}

\subsubsection{BaTiO$_3$ membrane}

Our motivation for studying a thin perovskite membrane as a
showcase for our method stems from the recent surge of 
interest in such systems.
This rapidly growing area of research has been fueled by the 
experimental breakthroughs in the preparation of unsupported 
oxide films via sacrifical layers.~\cite{lu-16}
The main advantage resides in the unprecedented possibility
of studying the impact of reduced size on the properties 
of perovskite crystals, and on the unprecedented degree of control
over the mechanical boundary conditions that a membrane geometry allows.~\cite{Hong-20}
The theoretical study of the phonon spectrum, a mainstay of the
current understanding of 3D complex oxides, provides a unique
view on the effects of dimensionality on, e.g., the stability
of the lattice against a ferroelectric distortion.
We shall provide a practical demonstration in the following.

Fig.~\ref{Fig_BTO} shows the dispersion of the optical phonons as obtained 
from the three different interpolation methods that we introduced in the
previous paragraphs. 
We have used an optimal value of $L$=12.0 bohr,
once again extracted by minimizing Eq.~(\ref{dl}) as
shown in Fig.~\ref{Fig_IFCs}.
Many of the trends that we have already observed for BN and SnS$_2$ crystals are
manifestly present: i) the highly dispersive LO modes are most affected by the 
improvements brought about by our new formalism, concretely by the enhanced treatment 
of screening; ii) the ZO branches exhibit a linear dispersion in long-wavelength 
limit, requiring explicit treatment of the out-of-plane DD interactions for
its qualitatively correct representation;
iii) the effect of the dynamical quadrupoles is minimal, and only barely appreciable 
in the dispersion of the second-highest ZO branch. 
Interestingly, our interpolation scheme results in an improved description of
selected \emph{transverse} optical branches as well.
We believe that the discrepancies produced by the existing scheme might be a
``collateral damage'' of its inaccurate description of the LO branches:
our 2D BaTiO$_3$ crystal appears to be a case where the small-${\bf q}$ dip in the 
dispersion of some LO modes is particularly pronounced, possibly affecting the corresponding 
TO branches as well.

The physical origin of this rather extreme behavior resides in the ferroelectric
low-energy mode of BaTiO$_3$, which is characterized by an abnormally large dipolar 
strength, $Z_l$.
(Recall that the linear dispersion coefficient of the dynamical matrix eigenvalues close to $\Gamma$
is proportional to the \emph{square} of $Z_l$.~\cite{sohier-17})
Interestingly, the minimal thickness of the film prevents this mode from going ``soft''
at any point in the 2D Brillouin zone (our centrosymmetric structure is, therefore, 
at least a metastable configuration of the crystal), pointing to a complete suppression 
of ferroelectricity in the ultrathin limit.
Studying the crossover between 2D and 3D physics as a function of slab
thickness in this system will be an
exciting topic for future studies. In this perspective, we expect the virtues 
of our interpolation method to become even more manifest as thickness increases.
Indeed, the near-surface dynamical quadrupoles in our scheme grow linearly with 
thickness because of the dipolar contribution in Eq.~(\ref{qtauz}), and eventually
might become crucially important for a qualitatively correct interpolation.

\section{Conclusions}

In summary, we have developed a rigorous analytical 
description of the long-range electrostatic screening and
interatomic forces in two-dimensional crystals, 
within a fundamental first-principles context.
As a first application, we have used it to develop 
an explicit formula, exact up to the quadrupolar
order, for the long-range part of the interatomic 
force constants.
Numerical tests on selected materials demonstrate its
superior accuracy in the interpolation of the
phonon bands, at no extra cost compared to 
the existing schemes.

Our formalism provides a general 
platform for treating long-range electrostatics in 2D 
systems, with an applicability that goes well beyond the
specifics of lattice dynamics.
First, one could use Eq.~(\ref{eph}) to reconstruct
the nonanalytic contributions to the scattering
potential in electron-phonon calculations, in a
similar spirit as in Refs.~\onlinecite{brunin-prl20,brunin-prb20,Jhalani-20,Park-20}.
Second, the results of Sec.~\ref{sec:hyperbolic} 
should allow for a natural incorporation of our formalism
into dielectric models of layered systems, e.g.,
in combination with the methods of Refs.~\cite{andersen-15,mohn-18,sohier-21,sponza-20}.
The exact 2D representation of the macroscopic dielectric function 
of an arbitrarily thick layer makes our approach particularly appealing
in this context, as it does not require any approximation  
(e.g., to the monopolar/dipolar interactions~\cite{andersen-15,sohier-21}),
or limiting assumption (e.g., about the separable character of the
ground-state wave functions~\cite{sponza-20}).
In turn, our exact treatment of higher-order
order multipolar couplings could facilitate the 
description and modeling of advanced electromechanical 
effects, such as flexoelectricity.~\cite{springolo-20}
Also, one could generalize the calculation of the
2D polarizability functions, $\chi({\bf q})$, to 
finite frequencies, and thereby facilitate the
use of modern many-body perturbation techniques
in low-dimensional systems.~\cite{freysoldt-08,Louie-20}
In a materials context, our methods appear 
well suited to treating emergent systems
that lie at the crossover between 2D and 3D, such as
oxide membranes, which are attracting a rapidly growing 
experimental interest.
Finally, generalizing our approach to one-dimensional 
nanowires could be another exciting topic for follow-up 
studies.

\begin{acknowledgments}
 We acknowledge the support of Ministerio de Economia,
 Industria y Competitividad (MINECO-Spain) through
 Grants  No.  MAT2016-77100-C2-2-P, No. PID2019-108573GB-C22  
 and Severo Ochoa FUNFUTURE center of excellence (CEX2019-000917-S);
 and of Generalitat de Catalunya (Grant No. 2017 SGR1506).
 This project has received funding from the European
 Research Council (ERC) under the European Union's
 Horizon 2020 research and innovation program (Grant
 Agreement No. 724529). Part of the calculations were performed at
 the Supercomputing Center of Galicia (CESGA).
\end{acknowledgments}

\appendix

\section{Supporting analytical derivations}

\label{app:derivations}

 {\em Proof of Eq.~(\ref{w_decomp}).} 
By using the results and definitions of Sec.~\ref{sec:screen}, and in particular by recalling that $\chisr \epsilon_{\rm lr}^{-1}= \chi$, 
we find
\begin{equation}
\begin{split}
(\underbrace{1 +  \nusr \chisr}_{\epsilon_{\rm sr}^{-1}}) & \wlr  \underbrace{(1 + \chisr \nusr)}_{(\epsilon_{\rm sr}^{-1})^\dagger}  = \\
   =&  (\epsilon_{\rm lr}^{-1} + \nusr \chisr \epsilon_{\rm lr}^{-1}) \nulr (1 + \chisr \nusr) \\
                                            =& (1 + \nulr \chi + \nusr \chi ) \nulr (1 + \chisr \nusr) \\
                                            =& (1 + \nu \chi  ) \nulr (1 + \chisr \nusr) \\
                                            =& \underbrace{(1 + \nu \chi  ) \nu}_W - (1 + \nu \chi  ) \nusr + (1 + \nu \chi  ) \nulr \chisr \nusr.
%
\end{split}
\end{equation}
We are left to show that the second and third terms on the rhs sum up to $-\wsr$, 
\begin{equation}
\begin{split}
 (1 + \nu \chi  ) \nusr - (1 + \nu \chi  )& \nulr \chisr \nusr = \\
    =&  (1 + \nu \chi  ) ( \underbrace{1 - \nulr \chisr}_{\epsilon_{\rm lr}}) \nusr \\
                                                               =&  \epsilon_{\rm lr} \nusr + \nu \chisr \nusr \\
                                                               =& \nusr - \nulr \chisr \nusr +  \nu \chisr \nusr \\
                                                               =& \wsr.
\end{split}
\end{equation}

\emph{Plane-wave representation of Eq.~(\ref{phiscr})}.
We shall work with the cell-periodic part of functions and operators at
a certain wavevector ${\bf q}$ in the Brillouin zone. 
We shall set the normalization conventions for the forward Fourier transform in 3D as
\begin{equation}
f({\bf G}) = \frac{1}{\sqrt{\Omega}} \int_{\rm cell} d^3 r e^{-i{\bf G \cdot r}} f({\bf r}),
\end{equation}
where $f$ is a generic cell-periodic function (not to be confused with the
range-separation function defined in the main text), and ${\bf G}$ belongs to
the reciprocal-space Bravais lattice of the crystal.
In other words, we use a basis for our full-space operators of the type
\begin{equation}
\langle {\bf r}| {\bf G}\rangle = \frac{1}{\sqrt{\Omega}} e^{i{\bf G \cdot r}}.
\end{equation}
On such a basis, the external charge of Eq.~(\ref{phiscr}) reads as
\begin{equation}
\langle {\bf G}|\rho^{\rm ext}_{\kappa \alpha}   \rangle = 
  -i \frac{Z_\kappa}{\sqrt{\Omega}} ({\bf G + q})_\alpha e^{i{\bf G} \cdot \bm{\tau}_\kappa },
\end{equation}
and the bare Coulomb kernel is 
\begin{equation}
W({\bf G+q},{\bf G'+q}) = \delta_{\bf G G'} \frac{4\pi}{|{\bf G + q}|^2}.
\end{equation}
With these definitions, our Eq.~(\ref{phiscr}) coincides with Eq.~(4.5)
of PCM.

In two dimensions, we use a mixed representation where the in-plane components
are treated in reciprocal space, while the out-of-plane coordinate is 
treated in real space.
The Fourier transform then reads as
\begin{equation}
\label{fou2D}
f({\bf G}_\parallel,z) = \frac{1}{\sqrt{S}} \int_{\rm cell} dx dy e^{-i{\bf G_\parallel \cdot r}} f({\bf r}),
\end{equation}
where $S$ is the cell surface.


 {\em Proof of Eq.~(\ref{nulr1}).}
By combining Eq.~(\ref{decomp_nu}) and Eq.~(\ref{yukawa}), 
our definition of $\nulr=\nu-\nusr$ reads as
\begin{equation}
\nulr({\bf K}_\parallel,z-z') = -\frac{2 \pi}{K_\parallel} \sum_{n\neq 0} (-1)^n  e^{-K_\parallel|z-z'-nL|}.
\end{equation}
Based on the assumption that $|z-z'|<L$, this can be written as
\begin{equation}
\begin{split}
\nulr({\bf K}_\parallel,z-z') = & -\frac{2 \pi}{K_\parallel} \left( e^{K_\parallel|z-z'|} + e^{-K_\parallel|z-z'|}  \right) \\
 &  \times \sum_{n=1}^{+\infty} (-1)^n e^{-n K_\parallel L}. 
\end{split} 
\end{equation}
The sum can be simplified by using the formula for
the geometric power series, converging for $|x|<1$,
\begin{equation}
\frac{1}{1-x} = 1 + x + x^2 + x^3 + \cdots
\end{equation}
We arrive at
\begin{equation}
\nulr({\bf K}_\parallel,z-z') = \frac{2 \pi}{K_\parallel} \cosh [K_\parallel(z-z')] \frac{2 e^{-K_\parallel L}}{1+e^{-K_\parallel L}},
\end{equation}
which coincides with Eq.~(\ref{nulr1}).

\section{Dipoles, quadrupoles and polarizabilities in 2D \label{3Dto2D}}

\label{app:multipoles}

In the following, we shall discuss how the physical quantities entering 
Eq.(\ref{rho_approx}) and Eq.~(\ref{chi_approx}) are related to the Born 
dynamical charges ($Z^{(\alpha)}_{\kappa \beta}$), 
dynamical quadrupoles ($Q^{(\alpha\gamma)}_{\kappa \beta}$) and 
macroscopic clamped-ion dielectric tensor ($\epsilon_{\alpha\beta}$) that
are calculated via standard linear-response techniques~\cite{gonze/lee,Royo2019} 
in a supercell geometry. 
[The macroscopic dielectric \emph{tensor} of a 3D crystal, $\epsilon_{\alpha\beta}$, 
should not be confused with the small-space dielectric \emph{function}, $\telr({\bf q})$,
that we define and use in the main text.]
There are two main differences that one needs to take into account:
(i) the electrical boundary conditions (EBC) are not the same, since the 
quantities entering Eq.~(\ref{chi_approx}) and Eq.~(\ref{rho_approx}) are 
intended to be calculated within the ``zone-boundary'' electrostatics, while the 
standard implementation of $Z^{(\alpha)}_{\kappa \beta}$, $Q^{(\alpha\gamma)}_{\kappa \beta}$ and 
$\epsilon_{\alpha\beta}$ assumes 3D short-circuit boundary conditions 
(as obtained by removing the nonanalytic ${\bf G}=0$ term from the Coulomb 
kernel); and (ii) the multipole moments are assumed to be taken with 
respect to the $z=0$ symmetry plane, rather than the unperturbed atomic 
location.

Regarding the Born charges and dielectric polarizabilities, one only needs
to worry about (i), since they are both dipolar in character and hence 
origin-independent.
The in-plane Born charges are unaltered by the EBC, i.e. for 
$\alpha=x,y$ we have
\begin{equation}
\hat{Z}^{(\alpha)}_{\kappa \beta} = Z^{(\alpha)}_{\kappa \beta}.
\end{equation}
Conversely, along the out-of-plane direction the so-called ``Callen charges'' must 
be used, consistently with the open-circuit EBC that the reference zone-boundary
electrostatics imposes along $z$,
\begin{equation}
\hat{Z}^{(z)}_{\kappa \beta} = \frac{Z^{(z)}_{\kappa \beta}}{\epsilon_{zz}}.
\end{equation}
The macroscopic polarizabilities of the 2D layer can be calculated via 
\begin{eqnarray}
\alpha^\parallel_{\alpha\beta} &=& \frac{L}{4\pi} (\epsilon_{\alpha\beta} - 1), \label{alp_par} \\
\alpha^\perp &=&  \frac{L}{4\pi}  (1 - \epsilon_{zz}^{-1}). \label{alp_perp}
\end{eqnarray}
(The indices $\alpha\beta$ run over the two in-plane components.)
Note that 
the parameters $\hat{\bf Z}_{\kappa\beta}$, $\bm{\alpha}^\parallel$ and
$\alpha^\perp$ are all independent of
the vacuum thickness (provided that the electron density of neighboring images
has negligible overlap), as required for well-defined materials properties.
The above results are consistent with the prescriptions of Refs.~\cite{sohier-16,sohier-17,tian-20}:
our work puts them on firmer theoretical grounds, by identifying them with the
exact limiting behavior of well-defined response functions.

Devising the conversion rules for the dynamical quadrupoles is slightly more
delicate, as different components mix up in a way that is not always intuitive.
Regarding the the mixed and out-of-plane components, one has 
\begin{subequations}
\label{qtauz}
\begin{align}
\hat{Q}^{(z \alpha)}_{\kappa\beta} &=  \frac{ Q^{(z \alpha)}_{\kappa\beta} + 
     \tau_{\kappa z} Z^{(\alpha)}_{\kappa\beta} } {\epsilon_{zz}}, \\
\hat{Q}^{(z z)}_{\kappa\beta}   &= \frac{  Q^{(zz)}_{\kappa\beta} + 2\tau_{\kappa z} Z^{(z)}_{\kappa\beta} } 
 { \epsilon_{zz}}. 
\end{align}
\end{subequations}
The dielectric constant at the denominator relates to the EBC change, 
analogously to the above discussion of the Born effective charges.
The addition of the Born effective charge times the $z$ coordinate of the atom
at the numerator, on the other hand,
takes care of the origin shift.
Indeed, the dynamical quadrupoles within DFPT can 
be written as a second moment of the charge density induced by an atomic 
displacement as~\cite{artlin}
\begin{equation}
Q^{(\alpha \gamma)}_{\kappa \beta} = \int d^3 r \, \rho_{\kappa \beta}({\bf r}) \, ({\bf r} - \bm{\tau}_\kappa)_\alpha
 ({\bf r} - \bm{\tau}_\kappa)_\gamma.
\end{equation} 
One can then break down the $z$-components of the round brackets as
\begin{equation}
({\bf r} - \bm{\tau}_\kappa)_z = z - {\tau}_{\kappa z},
\end{equation}
and after recalling that the Born charge can also be defined as a real-space moment,
\begin{equation}
Z^{(\alpha)}_{\kappa \beta} = \int d^3 r \, \rho_{\kappa \beta}({\bf r}) \, ({\bf r} - \bm{\tau}_\kappa)_\alpha,
\end{equation} 
one quickly arrives at Eq.~(\ref{qtauz}).
We are only left with working out the in-plane components, which can be readily converted as
\begin{equation}
\hat{Q}^{(\alpha \gamma)}_{\kappa\beta} = {Q}^{(\alpha \gamma)}_{\kappa\beta} - 4\pi \chi_{\alpha\gamma} \hat{Q}^{(z z)}_{\kappa\beta},
\end{equation}
where $\chi_{\alpha\beta}$ are the in-plane components of the macroscopic 
dielectric susceptibility tensor of the supercell.
It is interesting to note that the enforcement of the correct electrical
boundary conditions for a suspended 2D layer already endows the in-plane
quadrupoles with a contribution from the $(zz)$ component, $\hat{Q}^{(z z)}_{\kappa\beta}$.
Because of this, the traceless component that appears in Eq.~(\ref{rho_approx}) 
enjoys a particularly simple expression,
\begin{equation}
\label{trlq}
\hat{Q}^{(\alpha\gamma)}_{\kappa\beta} - \delta_{\alpha\gamma} \hat{Q}^{(z z)}_{\kappa\beta}= 
Q^{(\alpha\gamma)}_{\kappa\beta} - \epsilon_{\alpha\gamma} \hat{Q}^{(z z)}_{\kappa\beta}.
\end{equation}
One can show that all the components of $\hat{\bf Q}_{\kappa\beta}$ are all independent of the
vacuum thickness, $L$, unlike those of ${\bf Q}_{\kappa\beta}$.

Interestingly, a contribution of the out-of-plane dipoles to the electron-phonon
matrix elements involving the $A'_1$ branch of MoS$_2$ was recently identified in Ref.~\cite{deng2020}.
The above results nicely clarify the physical nature of the reported mechanism: the
out-of-plane dipoles contribute to $\hat{Q}^{(z z)}_{\kappa\beta}$ via Eq.~(\ref{qtauz})
and, in turn, to the longitudinal fields (mirror-even potentials) via Eq.~(\ref{trlq}).
(The $A'_1$ phonon is mirror-even, and hence cannot couple to an out-of-plane field.)
Thus, the mechanism of Ref.~\cite{deng2020} is understood, within our formalism, as a quadrupolar
contribution to the in-plane fields.
Note that, in addition to the aforementioned out-of-plane dipoles, our work reveals that
there are additional contributions to Eq.~(\ref{trlq}); their study will be an interesting topic
for future work.

\section{Relationship to the Coulomb cutoff technique}

\label{lautrec}

\begin{figure}
\begin{center}
\includegraphics[width=3.3in]{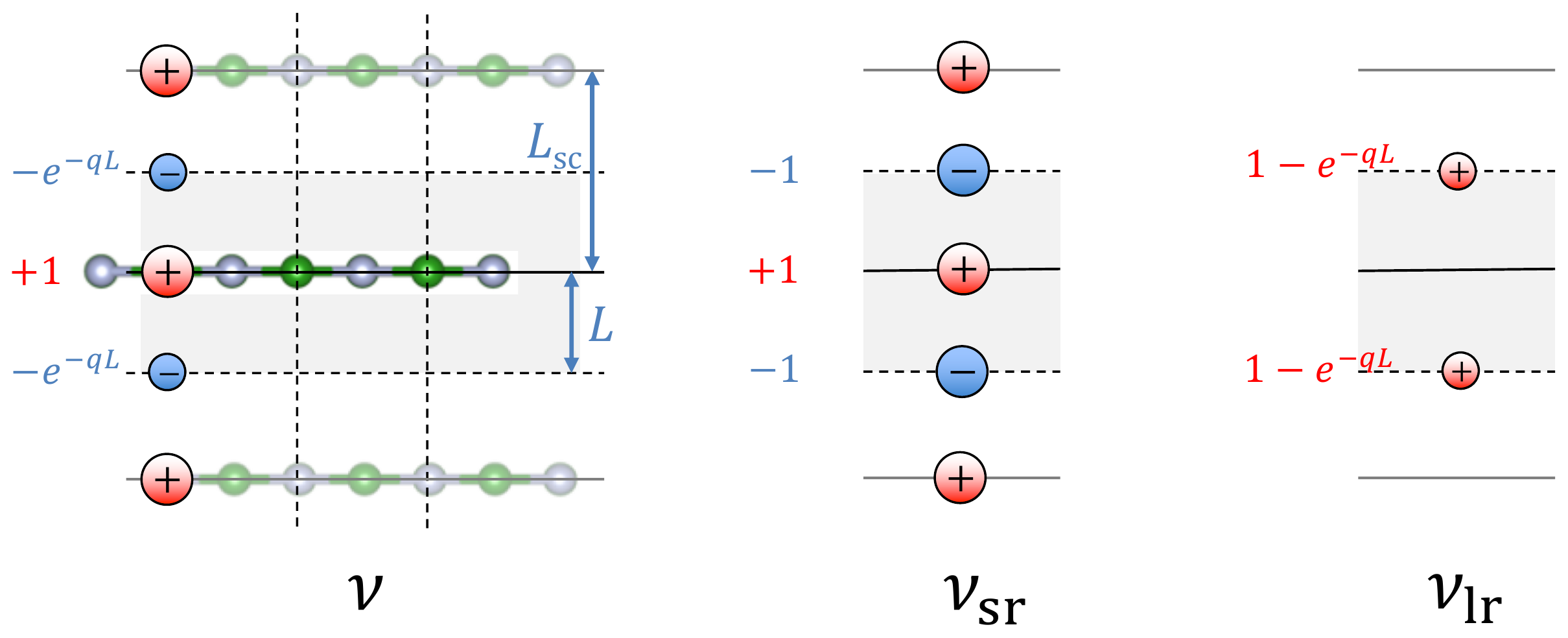}
\caption{Interpretation of the Coulomb truncation technique as an image-charge
method, and its separation into long-range and short-range contributions. The shaded areas
refer to the region of space where the three kernels exactly match
those illustrated in Fig.~\ref{kernel}. \label{imch}}
\end{center}
\end{figure}

The implementation of the Coulomb cutoff technique follows 
the prescriptions of Refs.~\onlinecite{sohrab-06,sohier-16,Sohier2017a}, and consists in
writing the open-boundary Coulomb kernel as
\begin{equation}
\nu ({\bf K_\parallel},G_n) = \frac{4\pi}{K_\parallel^2 + G_n^2} 
\left[ 1-e^{-K_\parallel L}\cos (G_n L) \right],
\end{equation}
where ${\bf K_\parallel}={\bf G}_\parallel + {\bf q}$ is an in-plane 
reciprocal-space vector (${\bf G}_\parallel$ spans the Bravais lattice of
the primitive 2D cell); $G_n$ form a discrete mesh along $z$ and 
$L=L_{\rm sc}/2$ is set to \emph{half} the supercell length in
the out-of-plane direction.
After observing that $G_n= \frac{\pi n}{L}$, we immediately obtain
the following expression for the macroscopic ${\bf G}_\parallel = 0$ component,
\begin{equation}
\nu ({\bf q},G_n) = \frac{4\pi}{q^2 + G_n^2} 
[ 1 - (-1)^n e^{-q L} ].
\end{equation}

To link these expressions to the arguments of Section~\ref{sec:kernel2d}, we shall
rewrite the prefactor in the square brackets as follows,
\begin{equation}
[ 1 - (-1)^n e^{-q L} ] = [ 1 - (-1)^n ] + (-1)^n (1 - e^{-q L})
\label{decomp}
\end{equation}
It is easy to see that the first term on the rhs corresponds to the
short-range ``zone-boundary'' electrostatics,
\begin{equation}
\label{nusrpw}
\nu_{\rm sr} ({\bf q},G_n) = \frac{4\pi}{q^2 + G_n^2} [ 1 - (-1)^n ].
\end{equation}
Indeed, the $[ 1 - (-1)^n ]$ prefactor can be regarded as an
implementation of the image-charge method illustrated in Fig.~\ref{kernel}.
(That this kernel is short-ranged is obvious from Eq.~(\ref{nusrpw}):
even values of the out-of-plane index $n$ are suppressed, thus excluding the
problematic $n=0$ term.)
This latter observation reveals that
the Coulomb cutoff technique can also be interpreted 
as an image-charge method: it only differs from $\nu_{\rm sr}$
in the prefactor $e^{-q L}$ that scales the negative images,
located at odd multiples of $L$ from the $z=0$ plane.
Then, we identify the long-range part of the kernel with the
remainder,
\begin{equation}
\nu_{\rm lr} ({\bf q},G_n) = 4\pi \frac{ (-1)^n}{q^2 + G_n^2} (1 - e^{-q L}).
\label{lr_G}
\end{equation}

To verify that Eq.~(\ref{lr_G}) is consistent with the formalism of the 
earlier sections, recall the following relation
for the Fourier series of the hyperbolic cosine function,
\begin{equation}
\int_{-\pi}^{\pi} \cosh(ax) \cos(nx) dx = (-1)^n \frac{2a \sinh(a\pi)}{a^2 + n^2}.
\end{equation}
By changing the variable to $z= Lx/\pi$, and by setting $q =\frac{a\pi}{L}$, 
we have
\begin{equation}
\int_{-L}^{L} \cosh(q z) \cos(G_n z) dz = (-1)^n \frac{2 q \sinh(q L)}{q^2 + G_n^2}.
\label{coshg}
\end{equation}
Then, observe that
\begin{equation}
\sinh(q L) \left[ 1 - \tanh \left(\frac{qL}{2} \right)  \right] = (1 - e^{-q L}).
\end{equation}
By combining the above, we eventually obtain
\begin{equation}
\int_{-L}^{L} \nu_{\rm lr} ({\bf q},z) e^{i G_n z} dz = \nu_{\rm lr} ({\bf q},G_n),
\end{equation}
with $\nu_{\rm lr} ({\bf q},z)$ defined as in Eq.~(\ref{nulr2}). 
The above formulas provide, therefore, the desired representation 
of the short-range and long-range Coulomb kernels in a supercell
context, together with 
an explicit reciprocal-space expression, Eq.~(\ref{coshg}),
for the hyperbolic cosine potential 
of Eq.~(\ref{potential}), which can be directly implemented in
a first-principles code. (Similar formulas can be easily derived
for the mirror-odd component.)

\section{Hyperbolic functions and traceless multipoles}

\label{app:traceless}

We shall provide a formal demonstration of our statement in Sec.\ref{sec:multipole},
that the hyperbolic functions consistently pick the traceless component of the first-order 
charge perturbation at any order in ${\bf q}$. 
To that end, we shall assume without loss of generality that 
the $x$ Cartesian axis is aligned with the propagation vector, ${\bf q}=(q,0,0)$,
and write the cell-periodic part of the external charge density perturbation, 
$\rho^{\rm sr,\bf q}_{\kappa \alpha}({\bf r})$, as a
lattice sum of the charge densities that are induced by a displacement 
of isolated atom,
\begin{equation}
\rho^{\rm sr,\bf q}_{\kappa \alpha}({\bf r}) = e^{-iqx} \sum_l \rho^{\rm sr}_{\kappa \alpha}({\bf r}-{\bf R}_{l\kappa}) e^{iqX_{l\kappa}}.
\end{equation}
The small-space representation of the charge response, $\tilde{\rho}^{\rm sr}$,
then reads
\begin{equation}
\begin{split}
\rho^{{\rm sr},(l)}(q) =& \frac{1}{S} \int_{\rm cell} dx dy  \int dz e^{-iq(x-X_{l\kappa)}} \varphi_l(z) \rho^{\rm sr}_{\kappa \alpha}({\bf r}-{\bf R}_{l\kappa}) \\
      =& \frac{1}{S} \int d^3r e^{-iqx} \varphi_l(z) \rho^{\rm sr}_{\kappa \alpha}({\bf r}),
\end{split}
\end{equation}
where the integral in the second line is taken over all space, and the origin is
set at the projection of the atom $\kappa$ of the $l=0$ cell on the $z=0$ plane, $(X_{0\kappa},Y_{0\kappa},0)$.
[Following the notation of the main text, $\varphi_l(z)$ stands for the hyperbolic cosine ($l=\parallel$) or
sine ($l=\perp$).]

\begin{table}
\setlength{\tabcolsep}{12pt}
\begin{center}
\begin{tabular}{c|cccc}
\hline \hline
     & 0 & 1     & 2 & 3 \\
\hline     
cosh ($\parallel$) & 1 & $-ix$  & $-x^2 + z^2$ &  $ix^3 -3ixz^2$ \\
sinh ($\perp$)  & 0 &  $z$    &  $-2ixz$     &  $3xz^2 - z^3$ \\
\hline
\hline
\end{tabular}
\caption{\label{tab:multi} Multipolar expansion of the basis functions $F^{\parallel, \perp}_n(x,z)$ 
in Cartesian coordinates. The columns correspond to $n$.}
\end{center}
\end{table}

The function in the integrand can be written as follows,
\begin{equation}
\label{cartesian}
e^{-iqx} \varphi_l(z) = \frac{ e^{q(-ix+z)} \pm e^{q(-ix-z)} }{2},
\end{equation}
where the plus and minus sign refer to cosine and sine, respectively.
The expansion of the exponential in powers of $q$ trivially leads to
\begin{equation}
e^{q(-ix+z)} \simeq 1 + q(-ix+z) + \frac{q^2}{2!} (-ix+z)^2 + \cdots
\end{equation}
If we write the complex number in the round brackets in terms of its
modulus, $r$, times a unitary phase, $e^{i\phi}$, we arrive at
\begin{equation}
e^{q(-ix+z)} = \sum_{n=0}^\infty \frac{q^n}{n!} r^n e^{i n\phi}.
\end{equation}
One can easily recognize the solutions of the Laplace equation 
in cylindrical coordinates, given by the $n$-th power of the radial coordinate $r$
times a cylindrical harmonic of the same order,
\begin{equation}
F_n(r,\phi) = r^n e^{i n\phi}.
\end{equation}

For any $n>0$, there are two (and only two) linearly independent solutions, which
we can write as
\begin{equation}
F^{\parallel, \perp}_n  (r,\phi) = \frac{ F_n(r,\phi) \pm F_n(r,-\phi)}{2}.
\end{equation}
(We have taken their mirror-even and mirror-odd linear combinations with respect to 
$z$-reflection.)
Finally, we have
\begin{equation}
e^{-iqx} \varphi_{\parallel, \perp}(z) = \sum_{n=0}^\infty \frac{q^n}{n!} F^{\parallel, \perp}_n  (r,\phi).
\end{equation}
This shows that the cosh and sinh basis functions correspond to the 2D Fourier 
transforms of $F^{\parallel, \perp}_n$;
it is easy to show that $e^{iqx} \varphi_l(z)$ are themselves solution of the
Laplace equation in two dimensions.

Based on the above, we can conclude that, at any given order $n>0$, 
there are two (and only two) independent multipolar 
component of the bounded charge distribution $\rho^{\rm sr}_{\kappa \alpha}({\bf r})$ that 
produce long-range electrostatic potentials; these are given by the 
integrals 
\begin{equation}
\label{muu}
M^{\parallel, \perp}_{\kappa \alpha}(n) = \frac{1}{S} \int d^3r F^{\parallel, \perp}_n (x,z) \rho^{\rm sr}_{\kappa \alpha}({\bf r}).
\end{equation}
From Eq.~(\ref{cartesian}) it is easy to work out a Cartesian representation
for the lowest orders, which we report in Table~\ref{tab:multi}.
This shows that the individual components of the Cartesian multipole tensors
[which are defined by replacing $F^{\parallel, \perp}_n (x,z)$ with $x^j z^k$ in
Eq.~(\ref{muu})] are not necessarily relevant for the long-range electrostatics -- only
their linear combinations, taken according to the prescriptions of Table~\ref{tab:multi},
are. 
These linear combinations result in removing the \emph{trace} of the Cartesian tensors at
any given order $j+k$ -- this is obvious in the $n=2$ case, where the mirror-even
quadrupole is given by the difference of the (diagonal) $x^2$ and $z^2$ components.
This is nicely consistent with Eq.~(\ref{rho_approx}).

\bibliography{merged,2d}

\end{document}